\documentclass[a4paper,10pt]{article}
\usepackage[cp1251]{inputenc}
\usepackage[english]{babel}
\usepackage{amsmath}
\usepackage{amssymb}
\usepackage{graphicx}
\usepackage[pdftex]{color}
\usepackage[square,numbers,comma,sort&compress]{natbib}
\usepackage[pdftex]{hyperref}
\hypersetup{
             final,
             colorlinks=true,
             linkcolor=blue,
             citecolor=blue
}

\DeclareMathOperator{\sgn}{sgn}

\begin{document}
%\vspace{-0.25cm}
\twocolumn[
\begin{center}
\Large{\bf Dielectric electron-hole liquid in monolayer heterostructures based on transition metal dichalcogenides}

\vspace{0.5cm}

\large{P.\,V. Ratnikov$^\dagger$}

\vspace{0.25cm}

\normalsize

\textit{Prokhorov General Physics Institute, Russian Academy of Sciences, ul. Vavilova 38, Moscow, 117942 Russia\\
$^\dagger$ratnikov@lpi.ru}
\end{center}

\vspace{0.25cm}
\begin{list}{}
{\rightmargin=0.85cm\leftmargin=0.85cm}
\item
\small{The possibility of the appearance of a dielectric electron-hole liquid (EHL) in monolayers of transition metal dichalcogenides and heterostructures based on them is considered. It is shown that the coherent pairing of electrons and holes in them leads to the formation of a dielectric EHL when the degree of circular polarization of the exciting light exceeds a certain threshold value. Below this value, a metallic EHL is realized. Some possible physical manifestations of the transition between these two types of EHL are noted.}

\end{list}\vspace{1cm}]

\begin{center}
\textbf{I. INTRODUCTION}
\end{center}

Over the past two decades, the physics of two-dimen-sional (2D) materials, in particular, transition metal dichalcogenides (TMD), has attracted the attention of many researchers. TMDs are layered materials and are described by the chemical formula $MX_2$, where $M$ is a transition metal atom (usually $M$=Mo,~W) and $X$ is a chalcogen atom ($X$=S,~ Se,~Te). Like the splitting of graphite into graphene layers \cite{RS2018} TMD can be split into thin films. The thinnest film consists of two layers of $X$ atoms, between which a layer of $M$ atoms is inserted. Such films are commonly referred to as TMD monolayers.

Features of the band structure make TMD monolayers attractive for use in valley-electronics \cite{Urbaszek2015}. In them, the valley-selective excitation of electron-hole pairs turns out to be possible, depending on the type of circular polarization of light: when right-handed light is absorbed, optical transitions occur in one valley, and for left-handed polarized light, in another~\cite{Durnev2018}.

The existence of a dielectric electron-hole liquid (EHL) was suggested in \cite{Keldysh1971}. Such a state can be realized due to the coherent pairing of electrons and holes.

In the first theoretical works devoted to the calculation of the energy of the EHL ground state (see, for example, the review \cite{Vashishta1974}), a free electron-hole gas was used as a zero approximation, the energy of which vanishes in the limit of zero density, and does not tend to exciton energy. This indicated that electron-hole correlations were not taken into account correctly in the region of low charge carrier densities.

It has long been known that the metallic state turns out to be unstable due to electron-hole correlations, which leads to the opening of an energy gap on the Fermi surface, the value of which at zero density coincides with the exciton binding energy \cite{Keldysh1964}. An estimate of the contribution of electron-hole correlations to the EHL energy with the introduction of a metal screening was made in \cite{Brinkman1973}. However, this approximation also looks unsatisfactory.

It was shown for the first time in \cite{Keldysh1975} that the coherent pairing of electrons and holes in three-dimensional (3D) semiconductors with isotropic bands leads to the formation of a dielectric EHL. The canonical transformation \cite{Keldysh1968} was used to construct the zero approximation of a system of electrons and holes interacting according to the Coulomb law.

In this work, we have studied the dielectric EHL in TMD monolayers and heterostructures based on them, taking into account the specifics of their band structure. We adapted the canonical transformation accordingly. A dielectric EHL can be more energetically favorable than a metallic one due to a decrease in the degeneracy factor of charge carriers.

\begin{center}
\textbf{II. MODEL FOUNDATIONS}
\end{center}

The Hamiltonian of a system of electrons and holes interacting according to the Coulomb law, taking into account the peculiarities of the band structure of TMD monolayers, has a form similar to the Hamiltonian in \cite{Keldysh1968}
\begin{equation}\label{Ham}
\begin{split}
\widehat{H}=&\sum_{\mathbf{p}s\tau}\left(\varepsilon^e_\mathbf{p}-\mu_e\right)a^\dag_{\mathbf{p}sK_\tau}a_{\mathbf{p}sK_\tau}+\\
&+\sum_{\substack{\mathbf{p}s\\ \tau=\sgn(s)}}\left(\varepsilon^h_\mathbf{p}-\mu_h\right)b^\dag_{\mathbf{p}sK_\tau}b_{\mathbf{p}sK_\tau}+\\
&+\frac{1}{2}\sum_{\substack{\mathbf{p}\mathbf{p}^\prime\mathbf{k}\\ ss^\prime\tau\tau^\prime}}V_\mathbf{k}
a^\dag_{\mathbf{p}sK_\tau}a^\dag_{\mathbf{p}^\prime s^\prime K_{\tau^\prime}}a_{\mathbf{p}^\prime+\mathbf{k}s^\prime K_{\tau^\prime}}a_{\mathbf{p}-\mathbf{k}sK_\tau}+\\
&+\frac{1}{2}\sum_{\substack{\mathbf{p}\mathbf{p}^\prime\mathbf{k}ss^\prime\\ \tau=\sgn(s)\\ \tau^\prime=\sgn(s^\prime)}}V_\mathbf{k}
b^\dag_{\mathbf{p}sK_\tau}b^\dag_{\mathbf{p}^\prime s^\prime K_{\tau^\prime}}b_{\mathbf{p}^\prime+\mathbf{k}s^\prime K_{\tau^\prime}}b_{\mathbf{p}-\mathbf{k}sK_\tau}-\\
&-\sum_{\substack{\mathbf{p}\mathbf{p}^\prime\mathbf{k}\\ ss^\prime\tau\\ \tau^\prime=\sgn(s^\prime)}}V_\mathbf{k}
a^\dag_{\mathbf{p}sK_\tau}b^\dag_{\mathbf{p}^\prime s^\prime K_{\tau^\prime}}b_{\mathbf{p}^\prime+\mathbf{k}s^\prime K_{\tau^\prime}}a_{\mathbf{p}-\mathbf{k}sK_\tau},
\end{split}
\end{equation}
where $a^\dag_{\mathbf{p}sK_\tau}$ ($a_{\mathbf{p}sK_\tau}$) and $b^\dag_{\mathbf{p}sK_\tau}$ ( $b_{\mathbf{p}sK_\tau}$) are operators of creation (annihilation) of electrons and holes with quasimomentum $\mathbf{p}$ and spin projection $s$ ($s=\pm\,^1 /_2$) in the valley of the Brillouin zone point $K_\tau$, $\tau=\pm$ is the valley index (for holes it coincides with the sign of the spin projection $\sgn(s)$, which is explicitly reflected in \eqref{Ham}); $\mu_{e(h)}$ --- chemical potential of electrons (holes) determined by the condition
\begin{equation}\label{Condition_for_mu}
\sum_{\mathbf{p}s\tau}\langle a^\dag_{\mathbf{p}sK_\tau}a_{\mathbf{p}sK_\tau}\rangle=
\sum_{\substack{\mathbf{p}s\\ \tau=\sgn(s)}}\langle b^\dag_{\mathbf{p}sK_\tau}b_{\mathbf{p}sK_\tau}\rangle=n,
\end{equation}
where $n$ --- 2D density of electrons and holes\footnote{In the case of unequal population of valleys, the 2D particle densities $n_+$ in the valley of point $K_+$ and $n_-$ in the valley of point $K_-$ ($n_++n_-=n$) should be introduced:
\begin{equation*}
\sum_{\mathbf{p}s}\langle a^\dag_{\mathbf{p}sK_\pm}a_{\mathbf{p}sK_\pm}\rangle=
\sum_{\mathbf{p}}\langle b^\dag_{\mathbf{p}\pm\frac{1}{2}K_\pm}b_{\mathbf{p}\pm\frac{1}{2}K_\pm}\rangle=n_\pm.
\end{equation*}
The unequal population of the valleys is achieved by excitation with light with a degree of circular polarization different from 0.5. If it tends to 1, the TMD monolayer behaves like a single-valley semiconductor.} $\langle~\rangle$ means averaging over the ground state.

The Coulomb interaction is chosen in the form of the Keldysh potential\footnote{Previously, for a metallic EHL, we adopted the usual 2D Coulomb potential, which gave very good agreement with experiment. In this case, the Keldysh potential was used to calculate excitons. To justify this choice, additional arguments were required (see our works \cite{Pekh2020, Pekh2021}). Here, the initial state (at $n\rightarrow0$) is a rarefied exciton gas, and the potential \eqref{KeldyshPotential} should be taken for it.}
\cite{Rytova1967, Keldysh1979}
\begin{equation}\label{KeldyshPotential}
V_\mathbf{k}=\frac{2\pi}{|\mathbf{k}|(1+r_0|\mathbf{k}|)}
\end{equation}
with the screening parameter $r_0$, which is determined by the best agreement between the calculated exciton binding energy in the limit of zero density and the experimentally measured one.

Charge carrier dispersion laws are
\begin{equation}
\varepsilon^e_\mathbf{p}=\frac{\mathbf{p}^2}{1+\sigma},~~\varepsilon^h_\mathbf{p}=\frac{\sigma\mathbf{p}^2}{1+\sigma},~~\sigma=\frac{m_e}{m_h}.
\end{equation}
We use a unit system with $e^2/\varepsilon_\text{eff}=\hbar=2m=1$, where $\varepsilon_\text{eff}=(\varepsilon_1+\varepsilon_2)/2$ is the effective static dielectric permittivity, determined by the dielectric permittivities of the media surrounding the TMD monolayer (for example, vacuum and substrate); $m=m_em_h/(m_e+m_h)$ is the reduced mass of an electron (with an effective mass $m_e$) and a hole (with an effective mass $m_h$). As before \cite{Pekh2021}, we assume $m_e$ and $m_h$ in the Hamiltonian \eqref{Ham} to be independent of $s$ and $\tau$.

The binding energy of a 2D exciton with the usual Coulomb interaction $2\pi e^2/\varepsilon_\text{eff}|\mathbf{k}|$ is taken as the unit of energy $E$ and temperature $T$
\begin{equation}
E_x=\frac{2me^4}{\hbar^2\varepsilon^2_\text{eff}},
\end{equation}
and as a unit of distance measurement we use its Bohr radius
\begin{equation}
a_x=\frac{\hbar^2\varepsilon_\text{eff}}{2me^2}.
\end{equation}
The 2D particle density $n$ is measured in units of $a^{-2}_x$. We assume the area of the system to be equal to unity.

We note some assumptions made when choosing the Hamiltonian \eqref{Ham}.

First, we assumed the processes of electron and hole scattering with spin flip to be suppressed due to the absence of the magnetic moment of the atoms constituting the crystal and magnetic impurities. However, it should be noted that spin flip processes can be resolved in TMD bilayers, which are composed of two monolayers, when charge carriers pass between the monolayers in a transverse electric field \cite{Gilardoni2021}.

The Hamiltonian \eqref{Ham} takes into account the specifics of the band structure of TMD monolayers. Recall that there is a large spin-orbit splitting of the valence band $\Delta_v\gtrsim100$ meV \cite{Wang2015}. The splitting of the conduction band is $\Delta_c=1-30$ meV \cite{Kosmider2013, Kormanyos2014}. The latter can be neglected at temperatures comparable to room temperature, assuming that the electrons are degenerate in spin. At an energy of exciting photons $\hbar\omega$ within $E_g<\hbar\omega<E_g+\Delta_v$ ($E_g$ is the band gap), holes are produced only on the upper spin branches of the valence band: with spin up in the valley of the point $K_+$ and with spin down in the valley of the point $K_-$ (see Fig.~\hyperlink{f1}{1}). Thus, summation over hole spin projections to \eqref{Ham} is equivalent to summation over valleys.

Secondly, we neglected the processes of intervalley transfer of charge carriers. The wave functions of carriers from different valleys are orthogonal, and the matrix elements corresponding to the intervalley transfer processes are small in comparison with the matrix elements left by us in the Hamiltonian \eqref{Ham} \cite{Lobaev1984}.

Third, we did not explicitly take into account the recombination of electrons and holes. Although it was taken into account indirectly in the choice of the ground state of the system of interacting electrons and holes in the form of a rarefied exciton gas. In this case, biexcitons and trions are lower in energy than excitons\footnote{For example, in WSe$_2$ monolayers encapsulated by boron nitride, the biexciton binding energy (i.e., the energy gain in the formation of a biexciton with respect to the energy of two excitons) is 17 meV \cite{Li2018}, which is $\simeq10\%$ of the exciton binding energy $E^{(exc)}_b=167\pm3$ meV \cite{Goryca2019}, while the intravalley electron trion binding energy ( i.e., the energy gain when an exciton captures an electron) is 35 meV \cite{Li2018}, which is $\simeq20\%$ of $E^{(exc)}_b$.}. However, due to the finite lifetime of all types of particles (both free carriers and composite ones), the number of biexcitons and trions is small compared to the number of excitons, since they do not have time to be formed from the latter in large numbers during short lifetimes. for relatively rare collisions, when $n\ll1$ (the time between two successive collisions can be on the order of the lifetime). These considerations are confirmed by the fact that the quenching of exciton lines in the photoluminescence spectrum of TMD monolayers occurs at a sufficiently high photoexcitation intensity and electron (hole) doping (the excess of the number of some carriers over the number of other carriers was up to $\sim10^{13}$ cm$^{-2}$) \cite{Mak2013, Ross2013, Yang2015}.

\begin{figure}[t!]
\begin{center}
\hypertarget{f1}{}
\includegraphics[width=0.45\textwidth]{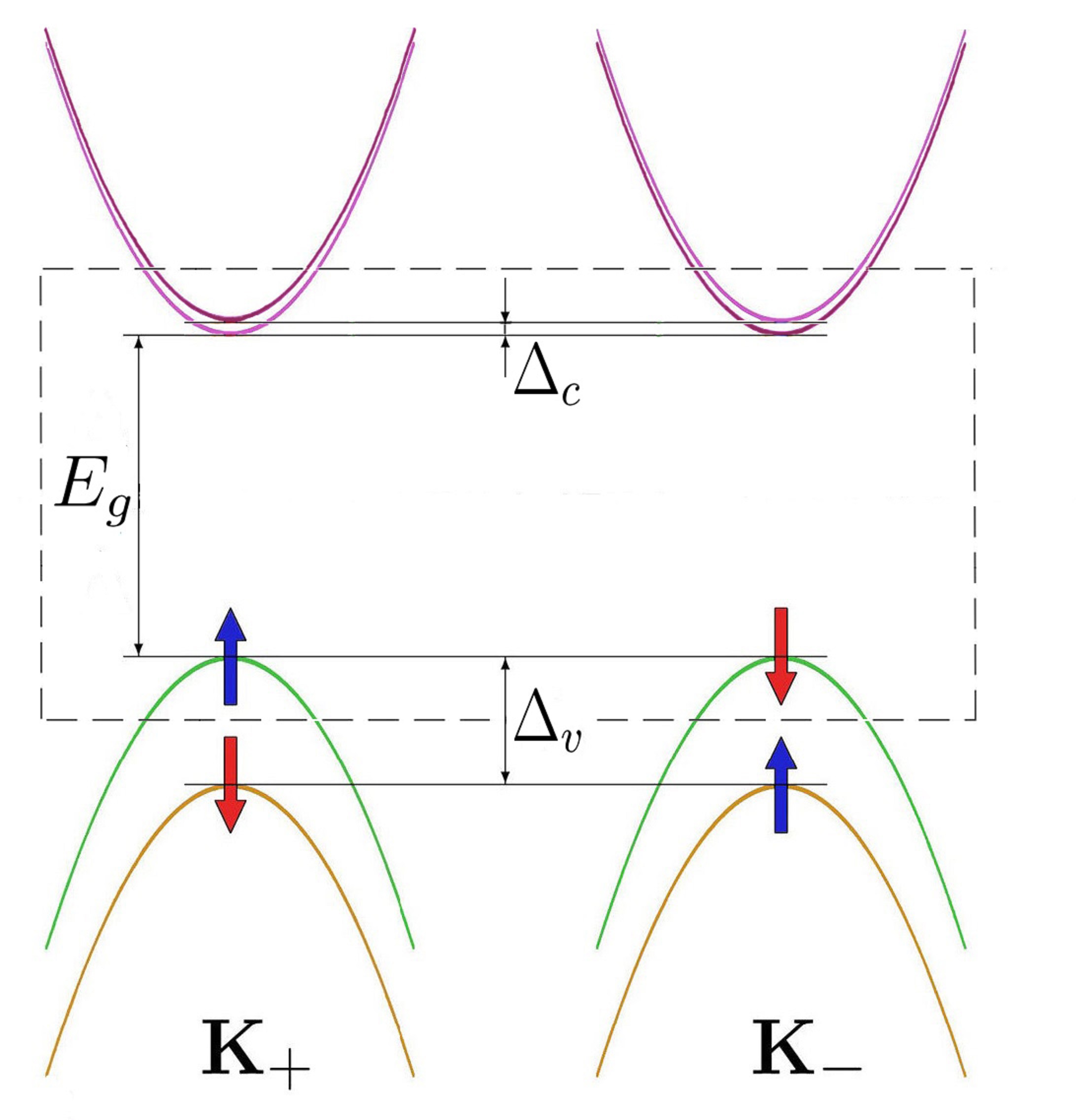}
\end{center}
FIG. 1. Band structure of TMD monolayers. The lower conduction band and the upper valence band are shown at two points $K_+$ and $K_-$. The arrows indicate the spin orientations of the valet band branches. The spin splitting of the valence band is equal to $\Delta_v$. In view of the small spin splitting in the $\Delta_c$ conduction band, the spin branches in it are distinguished by a tone: the lighter ones correspond to the spin up, the darker ones correspond to the spin down. The zone gap is equal to $E_g$. The dotted line marks the branches populated upon photoexcitation.
\end{figure}

On the other hand, the question of the stability of the ground state in the 2D case is qualitatively similar to that in the 3D case. In the density range $n_B\lesssim n\lesssim n_{dm}$ ($n_B\simeq10^{-3}$ --- the density at which the state constructed from biexcitons becomes unstable; $n_{dm}$ --- metal-dielectric transition density) the ground state of a system of electrons and holes interacting according to the Coulomb law is built from excitons \cite{Keldysh1971, Lobaev1988}.

In what follows, it is assumed that the density $n$ falls within the above density range (the density $n_{dm}$ for TMD monolayers was calculated by us in a previous paper \cite{Pekh2021}). The ground state built from trions is realized under conditions of electron (hole) doping (see also \cite{Mak2013, Ross2013, Yang2015}). In our case, there is no such doping.

Recombination is also qualitatively taken into account when a dynamic equilibrium between the number of electron-hole pairs produced in the regime of continuous photoexcitation and recombining particles is implied. This allows us to consider $n$ as a given value.

The band gap of TMD semiconductor monolayers $E_g\simeq2$ eV \cite{Durnev2018} is large compared to the characteristic energies (for example, the exciton binding energy does not exceed 0.4 eV \cite{Goryca2019, Eknapakul2014, He2014, Rigosi2016}), so we we use the one-zone approximation.

\begin{center}
\textbf{III. COHERENT PAIRING OF ELECTRONS AND HOLES}
\end{center}

As indicated above, the ground state of a system of electrons and holes interacting according to the Coulomb law is built from excitons. To take this situation into account, we make a canonical transformation \cite{Keldysh1968} of the Hamiltonian \eqref{Ham}
\begin{equation}\label{TransformationOperators}
\Lambda_\mathbf{p}=SL_\mathbf{p}S^\dag,
\end{equation}
where for convenience of notation the following columns of operators are introduced
\begin{equation*}
\begin{split}
L_\mathbf{p}&=\begin{pmatrix}
               A_{\mathbf{p}+} \\
               A_{\mathbf{p}-} \\
               B_\mathbf{p}
             \end{pmatrix},\\
                            A_{\mathbf{p}\tau}&=\begin{pmatrix}
                                                  a_{\mathbf{p}+\frac{1}{2}K_\tau} \\
                                                  a_{\mathbf{p}-\frac{1}{2}K_\tau}
                                                \end{pmatrix},~~B_\mathbf{p}=\begin{pmatrix}
                                                                               b^\dag_{-\mathbf{p}+\frac{1}{2}K_+} \\
                                                                               b^\dag_{-\mathbf{p}-\frac{1}{2}K_-}
                                                                             \end{pmatrix}.
\end{split}
\end{equation*}
The $\Lambda_\mathbf{p}$ column is composed of new Fermi operators, which, to distinguish them from the old operators, are denoted by the corresponding Greek letters with the same indices and in the same order as in the $L_\mathbf{p}$ column.

In the case of equal population of valleys\footnote{The case of unequal population of valleys is considered separately below.} The unitary operator $S$ is defined as
\begin{equation}\label{S_operator}
S=\exp\left\{\frac{i}{\sqrt{2}}\sum_{\mathbf{p}\tau}L^\dag_{\mathbf{p}\tau}\widehat{F}_\mathbf{p}L_{\mathbf{p}\tau}\right\},
\end{equation}
where
\begin{equation*}
\begin{split}
L_{\mathbf{p}\tau}&=\begin{pmatrix}
                     A_{\mathbf{p}\tau} \\
                     B_\mathbf{p}
                   \end{pmatrix},\\
                   \widehat{F}_\mathbf{p}&=\begin{pmatrix}
                                                            O & \widehat{\Phi}_\mathbf{p} \\
                                                            \widehat{\Phi}^\dag_\mathbf{p} & O
                                                          \end{pmatrix},~~\widehat{\Phi}_\mathbf{p}=-i\begin{pmatrix}
                                                                                                        \delta_\mathbf{p} & \gamma_\mathbf{p} \\
                                                                                                        \gamma_\mathbf{p} & \delta_\mathbf{p}
                                                                                                      \end{pmatrix},
\end{split}
\end{equation*}
$\gamma_\mathbf{p}$ and $\delta_\mathbf{p}$ are quasi-momentum functions determined from the condition of minimum energy and stability of the ground state of the system.

By direct calculations we find
\begin{equation}
\Lambda_\mathbf{p}=\widehat{R}_\mathbf{p}L_\mathbf{p},
\end{equation}
where the matrix $6\times6$
\begin{equation*}
\widehat{R}_\mathbf{p}=\begin{pmatrix}
                         \frac{1}{2}\left(M_\mathbf{p}+I\right) & \frac{1}{2}\left(M_\mathbf{p}-I\right) & N_\mathbf{p} \\
                         \frac{1}{2}\left(M_\mathbf{p}-I\right) & \frac{1}{2}\left(M_\mathbf{p}+I\right) & N_\mathbf{p} \\
                         -N_\mathbf{p} & -N_\mathbf{p} & M_\mathbf{p}
                       \end{pmatrix}
\end{equation*}
is given by the $2\times2$ matrices ($I$ is the identity matrix)
\begin{equation*}
M_\mathbf{p}=\begin{pmatrix}
               \cos\gamma_\mathbf{p}\cos\delta_\mathbf{p} & -\sin\gamma_\mathbf{p}\sin\delta_\mathbf{p} \\
              -\sin\gamma_\mathbf{p}\sin\delta_\mathbf{p} &  \cos\gamma_\mathbf{p}\cos\delta_\mathbf{p}
             \end{pmatrix},
\end{equation*}
\begin{equation*}
N_\mathbf{p}=\frac{1}{\sqrt{2}}\begin{pmatrix}
               \cos\gamma_\mathbf{p}\sin\delta_\mathbf{p} & \sin\gamma_\mathbf{p}\cos\delta_\mathbf{p} \\
               \sin\gamma_\mathbf{p}\cos\delta_\mathbf{p} & \cos\gamma_\mathbf{p}\sin\delta_\mathbf{p}
             \end{pmatrix}.
\end{equation*}
Two types of pairing are possible: singlet $\gamma_\mathbf{p}=\varphi_\mathbf{p}$ and $\delta_\mathbf{p}\equiv0$ (total electron and hole spin $\widetilde{S}=0$) or the triplet $\gamma_\mathbf{p}\equiv0$ and $\delta_\mathbf{p}=\varphi_\mathbf{p}$ ($\widetilde{S}=1$). Operators take the form

\begin{equation}\label{NewOperators}
\begin{split}
&\alpha_{\mathbf{p}sK_\tau}=\\
&=\frac{1}{2}\left(\cos\varphi_\mathbf{p}+1\right)a_{\mathbf{p}sK_\tau}+\frac{1}{2}\left(\cos\varphi_\mathbf{p}-1\right)a_{\mathbf{p}sK_{-\tau}}+\\
&+\frac{1}{\sqrt{2}}\sin\varphi_\mathbf{p}\left(\delta_{\widetilde{S}0}b^\dag_{-\mathbf{p}-sK_{-\sgn(s)}}+\delta_{\widetilde{S}1}b^\dag_{-\mathbf{p}sK_{\sgn(s)}}\right),\\
&\beta_{\mathbf{p}sK_{\sgn(s)}}=\cos\varphi_\mathbf{p}b_{\mathbf{p}sK_{\sgn(s)}}-\\
&-\frac{1}{\sqrt{2}}\sin\varphi_\mathbf{p}
\left[\delta_{\widetilde{S}0}\left(a^\dag_{-\mathbf{p}-sK_+}+a^\dag_{-\mathbf{p}-sK_-}\right)+\right.\\
&\left.+\delta_{\widetilde{S}1}\left(a^\dag_{-\mathbf{p}sK_+}+a^\dag_{-\mathbf{p}sK_-}\right)\right].
\end{split}
\end{equation}
The matrix $\widehat{R}_\mathbf{p}$ is, as it should be, a rotation matrix. In particular, $\det\widehat{R}_\mathbf{p}\equiv1$ is obtained by direct calculation. To show what kind of rotation it performs for a particular type of pairing, we introduce the columns
\begin{equation*}
\begin{split}
\widetilde{L}_{\mathbf{p}s}&=\begin{pmatrix}
                              \widetilde{L}_{\mathbf{p}s+} \\
                              \widetilde{L}_{\mathbf{p}s-}
                            \end{pmatrix},~~\widetilde{L}_{\mathbf{p}s\pm}=\begin{pmatrix}
                                                                             a_{\mathbf{p}sK_+} \\
                                                                             a_{\mathbf{p}sK_-} \\
                                                                             b^\dag_{-\mathbf{p}\pm sK_{\pm\sgn(s)}}
                                                                           \end{pmatrix},\\
\widetilde{\Lambda}_{\mathbf{p}s}&=\begin{pmatrix}
                              \widetilde{\Lambda}_{\mathbf{p}s+} \\
                              \widetilde{\Lambda}_{\mathbf{p}s-}
                            \end{pmatrix},~~\widetilde{\Lambda}_{\mathbf{p}s\pm}=\begin{pmatrix}
                                                                             \alpha_{\mathbf{p}sK_+} \\
                                                                             \alpha_{\mathbf{p}sK_-} \\
                                                                             \beta^\dag_{-\mathbf{p}\pm sK_{\pm\sgn(s)}}
                                                                           \end{pmatrix}.
\end{split}
\end{equation*}
The sign ``$+$'' in $\widetilde{L}_{\mathbf{p}s}$ and $\widetilde{\Lambda}_{\mathbf{p}s}$ corresponds to triplet pairing, the sign ``$-$'' --- singlet pairing. Then the transformation \eqref{NewOperators} will be rewritten
\begin{equation*}
\widetilde{\Lambda}_{\mathbf{p}s}=\widehat{R}^\prime_\mathbf{p}\widetilde{L}_{\mathbf{p}s},~~\widehat{R}^\prime_\mathbf{p}=
\begin{pmatrix}
  \mathcal{R}_\mathbf{p} & O \\
  O & \mathcal{R}_\mathbf{p}
\end{pmatrix},
\end{equation*}
where the matrix
\begin{equation*}
\mathcal{R}_\mathbf{p}=\begin{pmatrix}
                         \frac{1}{2}\left(\cos\varphi_\mathbf{p}+1\right) & \frac{1}{2}\left(\cos\varphi_\mathbf{p}-1\right) & \frac{1}{\sqrt{2}}\sin\varphi_\mathbf{p} \\
                         \frac{1}{2}\left(\cos\varphi_\mathbf{p}-1\right) & \frac{1}{2}\left(\cos\varphi_\mathbf{p}+1\right) & \frac{1}{\sqrt{2}}\sin\varphi_\mathbf{p} \\
                         -\frac{1}{\sqrt{2}}\sin\varphi_\mathbf{p}        & -\frac{1}{\sqrt{2}}\sin\varphi_\mathbf{p}        & \cos\varphi_\mathbf{p}
                       \end{pmatrix}
\end{equation*}
is the rotation matrix in 3D space through the angle $\varphi_\mathbf{p}$ around an axis lying in the $xy$ plane at an angle $-\hspace{-0.05cm}\,^\pi\hspace{-0.05cm}/ _4$ to the $x$ axis. The matrix $\widehat{R}^\prime_\mathbf{p}$ differs from the matrix $\widehat{R}_\mathbf{p}$ only by an even permutation of rows and columns.

The Hamiltonian \eqref{Ham} after the transformation \eqref{TransformationOperators} takes the form \cite{Keldysh1968, Lobaev1988}\footnote{Skipping going forward, we point out that the equality \eqref{Norm_phi2} was used to highlight the last term in \eqref{NewHam}.}
\begin{equation}\label{NewHam}
\widehat{\mathcal{H}}=S\widehat{H}S^\dag=\widetilde{U}\left\{\varphi_\mathbf{p}\right\}+\widehat{\mathcal{H}}_0+\widehat{\mathcal{H}}_i-\mu n,
\end{equation}
where $\mu=\mu_e+\mu_h$, $\widetilde{U}\left\{\varphi_\mathbf{p}\right\}$ is the numerical functional arising from normalizing the Hamiltonian
\begin{equation}\label{Ut}
\begin{split}
\widetilde{U}\left\{\varphi_\mathbf{p}\right\}=&2\sum_\mathbf{p}\varepsilon_\mathbf{p}\sin^2\varphi_\mathbf{p}-\\
-&2\sum_{\mathbf{p}\mathbf{p}^\prime}V_{\mathbf{p}-\mathbf{p}^\prime}\left(\sin^2\varphi_\mathbf{p}\sin^2\varphi_{\mathbf{p}^\prime}+\right.\\
&\left.+\cos\varphi_\mathbf{p}\sin\varphi_\mathbf{p}\cos\varphi_{\mathbf{p}^\prime}\sin\varphi_{\mathbf{p}^\prime}\right),
\end{split}
\end{equation}
where $\varepsilon_\mathbf{p}=\varepsilon^e_\mathbf{p}+\varepsilon^h_\mathbf{p}$. The twos are due to summation over $s$.

The operators $\widehat{\mathcal{H}}_0$ and $\widehat{\mathcal{H}}_i$ are given in Appendix A.

The density of new quasiparticles should be determined in such a way as the density of the original quasiparticles \eqref{Condition_for_mu}
\begin{equation}\label{Norm_n}
\sum_{\mathbf{p}s\tau}\langle\alpha^\dag_{\mathbf{p}sK_\tau}\alpha_{\mathbf{p}sK_\tau}\rangle=
\sum_{\mathbf{p}s}\langle\beta^\dag_{\mathbf{p}sK_{\sgn(s)}}\beta_{\mathbf{p}sK_{\sgn(s)}}\rangle=n.
\end{equation}

Substituting the expressions of the new operators \eqref{NewOperators} into \eqref{Norm_n} and taking the half-sum of both sums into \eqref{Norm_n}, we find
\begin{equation}\label{Norm_n_neue}
\begin{split}
&\sum_{\mathbf{p}s\tau}\left[\frac{1}{2}\cos^2\varphi_\mathbf{p}\langle a^\dag_{\mathbf{p}sK_\tau}a_{\mathbf{p}sK_\tau}\rangle-\right.\\
&-\frac{1}{2}\sin^2\varphi_\mathbf{p}\langle a^\dag_{\mathbf{p}sK_\tau}a_{\mathbf{p}sK_{-\tau}}\rangle+\frac{1}{\sqrt{2}}\cos\varphi_\mathbf{p}\sin\varphi_\mathbf{p}\times\\
&\times\left(\delta_{\widetilde{S}0}\langle a^\dag_{\mathbf{p}sK_\tau}b^\dag_{-\mathbf{p}-sK_{-\sgn(s)}}+
b_{-\mathbf{p}-sK_{-\sgn(s)}}a_{\mathbf{p}sK_\tau}\rangle+\right.\\
&\left.+\delta_{\widetilde{S}1}\langle a^\dag_{\mathbf{p}sK_\tau}b^\dag_{-\mathbf{p}sK_{\sgn(s)}}+
b_{-\mathbf{p}sK_{\sgn(s)}}a_{\mathbf{p}sK_\tau}\rangle\right)+\\
&\left.+\frac{1}{4}\cos2\varphi_\mathbf{p}\langle b^\dag_{\mathbf{p}sK_{\sgn(s)}}b_{\mathbf{p}sK_{\sgn(s)}}\rangle+
\frac{1}{2}\sin^2\varphi_\mathbf{p}\right]=n.
\end{split}
\end{equation}

Mean $\langle a^\dag_{\mathbf{p}sK_\tau}a_{\mathbf{p}sK_\tau}\rangle$ and $\langle b^\dag_{\mathbf{p}sK_{\ sgn(s)}}b_{\mathbf{p}sK_{\sgn(s)}}\rangle$ are zero: all levels of single-particle Fermi excitations $|\mathbf{p}sK_\tau\rangle$ (electrons) and $|\mathbf{p}sK_{\sgn(s)}\rangle$ (holes) lie above $\mu_e$ and $\mu_h$, and the states are not occupied for new quasiparticles \cite{Keldysh1968}. The second and third terms in \eqref{Norm_n_neue} can also be set equal to zero, since we can take advantage of the arbitrariness in choosing the function $\varphi_\mathbf{p}$ and introduce a condition for it, similar to \cite{Keldysh1968}
\begin{equation}\label{Norm_phi1}
\begin{split}
&\langle a^\dag_{\mathbf{p}sK_\tau}a_{\mathbf{p}sK_{-\tau}}\rangle=\langle a^\dag_{\mathbf{p}sK_\tau}b^\dag_{-\mathbf{p}-sK_{-\sgn(s)}}\rangle=\\
&=\langle b_{-\mathbf{p}-sK_{-\sgn(s)}}a_{\mathbf{p}sK_\tau}\rangle=\langle a^\dag_{\mathbf{p}sK_\tau}b^\dag_{-\mathbf{p}sK_{\sgn(s)}}\rangle=\\
&=\langle b_{-\mathbf{p}sK_{\sgn(s)}}a_{\mathbf{p}sK_\tau}\rangle=0.
\end{split}
\end{equation}

Thus, we arrive at the equality
\begin{equation}\label{Norm_phi2}
2\sum_\mathbf{p}\sin^2\varphi_\mathbf{p}=n.
\end{equation}

When \eqref{Norm_phi1} is executed, the averages $\langle\widehat{\mathcal{H}}_0\rangle$ and $\langle\widehat{\mathcal{H}}_i\rangle$ are zero. This means that in the self-consistent approximation, the energy of the system is determined by minimizing the numerical functional \eqref{Ut}.

In order for the condition \eqref{Norm_phi2} to be taken into account automatically, we pass from summation over $\mathbf{p}$ to integration over $q=p/\widetilde{p}$ by analogy with the 3D case
\cite{Lobaev1988}
\begin{equation}\label{pt}
\widetilde{p}=\frac{1}{p_0}\sqrt{\frac{2\pi n}{\nu_e}},
\end{equation}
where
\begin{equation*}
p_0=\sqrt{\frac{2}{\nu_e}\int\limits_0^\infty\frac{qdq}{1+z^2_q}},~~z_q=\cot\varphi_q.
\end{equation*}

The energy $E_0$ per one electron-hole pair is found by minimizing the functional
\begin{equation}
\begin{split}
&E_0\left\{z_q\right\}=\frac{4}{\nu_er^2_sp^4_0}\int\limits_0^\infty\frac{q^3dq}{1+z^2_q}-\\
&-\frac{8\sqrt{2}}{\pi\nu_er_sp^3_0}\int\limits_0^\infty\frac{q^2dq}{1+z^2_q}\int\limits_0^1\frac{1+z_qz_{q\xi}}{1+z^2_{q\xi}}\times\\
&\times\left[K(\xi)-\frac{1}{1+\xi}\widetilde{K}\left(\frac{2\sqrt{\xi}}{1+\xi};\,\widetilde{r}_0q(1+\xi)\right)\right]\xi d\xi,
\end{split}
\end{equation}
where the average distance between electrons is introduced
\begin{equation*}
r_s=\sqrt{\frac{\nu_e}{\pi n}},
\end{equation*}
$K(k)$ is the complete elliptic integral of the first kind, we introduced the function
\begin{equation*}
\widetilde{K}(k;\,\rho)=\int\limits_0^{\pi/2}\frac{\rho dx}{\rho\sqrt{1-k^2\sin^2x}+1},
\end{equation*}
which in the limit $\rho\rightarrow\infty$ goes over to $K(k)$, and the notation $\widetilde{r}_0=r_0\widetilde{p}$.

We have chosen the functions
\begin{equation}\label{TrialFunctions}
z_q=A(1+q^2)^\alpha+B
\end{equation}
as the trial functions with variational parameters $A$ and $B$, $\alpha\approx2$ [usually $\alpha=1.94-1.97$].

The calculation of the correlation corrections associated with the multiple creation and annihilation of electron-hole pairs was carried out according to the Nozi\`{e}res--Pines method \cite{Keldysh1975, Combescot1972}. Previously, we used this method to calculate the correlation energy for a metallic EHL in heterostructures based on TMD monolayers \cite{Pekh2020, Pekh2021}. A significant difference between the current calculations is the use of the \eqref{KeldyshPotential} potential when also calculating the correlation contribution. Due to the cumbersomeness of the formulas and the absence of fundamentally new results, we do not present the corresponding expressions here. We also note that this contribution in the region of small $n$ turns out to be small compared to the exchange contribution (in absolute value). A typical dependence of $E_0$ (after minimizing the functional \eqref{Ut}) on $n$ is shown in Fig.~\hyperlink{f2}{2}.

\begin{figure}[t!]
\begin{center}
\hypertarget{f2}{}
\includegraphics[width=0.45\textwidth]{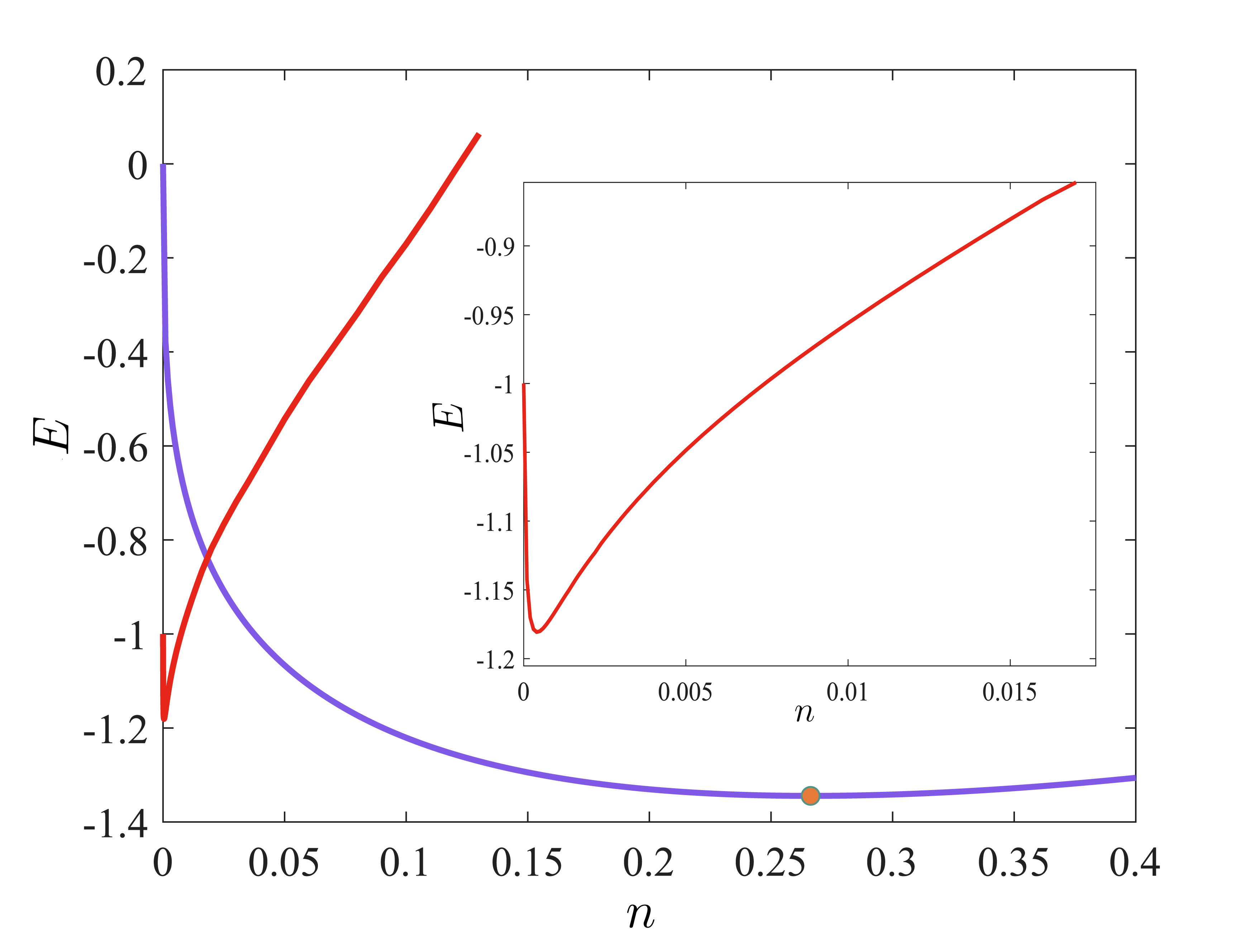}
\end{center}
FIG. 2. Dependence of the energy of the dielectric EHL (red curve) and metallic EHL (light blue curve) on the charge carrier density $n$ in the MoS$_2$ monolayer on the SiO$_2$ substrate in the case of equal population of the valleys. The yellow dot marks the minimum energy of the metallic EHL (its binding energy with a minus sign). The inset shows an enlarged section of the curve for the dielectric EHL.
\end{figure}

\begin{center}
\textbf{IV. UNEQUAL POPULATION OF THE VALLEYS}
\end{center}

In the unitary operator \eqref{S_operator} we must explicitly take into account the difference between the valleys at point $K_+$ (it is populated by electron-hole pairs with density $n_+$) and at point $K_-$ (density of $n_-$ pairs) . This means that in the expression in the \eqref{S_operator} exponent, we should select different functions for the coherent pairing of electrons and holes, when both of them are in the valley at the point $K_+$ or in the valley at the point $K_-$, or one particle is in one valley, and another particle is in another valley. For example, the member $\gamma_\mathbf{p}a^\dag_{\mathbf{p}-\frac{1}{2}K_+}b^\dag_{-\mathbf{p}+\frac{1} {2}K_+}$, which describes the process of the production of an electron-hole pair in the valley at the point $K_+$ with singlet pairing, one should compare the term $\gamma^{(+)}_\mathbf{p}a^\dag_ {\mathbf{p}-\frac{1}{2}K_+}b^\dag_{-\mathbf{p}+\frac{1}{2}K_+}$, and $\gamma_\mathbf{p}a^\dag_{\mathbf{p}-\frac{1}{2}K_-}b^\dag_{-\mathbf{p}+\frac{1}{2}K_-}$ (pair birth in the valley at point $K_-$) --- term $$\gamma^{(-)}_\mathbf{p}a^\dag_{\mathbf{p}\frac{1}{2}K_- }b^\dag_{-\mathbf{p}-\frac{1}{2}K_-};$$ in the case of intervalley pairing --- the term $$\widetilde{\gamma}_\mathbf{p}a^\dag_{\mathbf{p}+\frac{1}{2}K_+}b^\dag_{-\mathbf{p}-\frac{1}{2}K_-}$$ or $$\widetilde{\gamma}_\mathbf{p}a^\dag_{\mathbf{p}-\frac{1}{2}K_-}b^\dag_{-\mathbf{p}+\frac{1}{2}K_+}.$$ Similarly for triplet pairing
\begin{equation*}
\begin{split}
&\delta_\mathbf{p}a^\dag_{\mathbf{p}+\frac{1}{2}K_+}b^\dag_{-\mathbf{p}+\frac{1}{2}K_+}\,\rightarrow\,
\delta^{(+)}_\mathbf{p}a^\dag_{\mathbf{p}+\frac{1}{2}K_+}b^\dag_{-\mathbf{p}+\frac{1}{2}K_+},\\
&\delta_\mathbf{p}a^\dag_{\mathbf{p}-\frac{1}{2}K_-}b^\dag_{-\mathbf{p}-\frac{1}{2}K_-}\,\rightarrow\,
\delta^{(-)}_\mathbf{p}a^\dag_{\mathbf{p}-\frac{1}{2}K_-}b^\dag_{-\mathbf{p}-\frac{1}{2}K_-},\\
&\delta_\mathbf{p}a^\dag_{\mathbf{p}\pm\frac{1}{2}K_\mp}b^\dag_{-\mathbf{p}\pm\frac{1}{2}K_\pm}\,\rightarrow\,
\widetilde{\delta}_\mathbf{p}a^\dag_{\mathbf{p}\pm\frac{1}{2}K_\mp}b^\dag_{-\mathbf{p}\pm\frac{1}{2}K_pm}.
\end{split}
\end{equation*}

For pair destruction processes, substitutions are made in the same way.

According to these replacements, the matrix $\widehat{F}_\mathbf{p}$ in \eqref{S_operator} becomes dependent on the valley index:
\begin{equation*}
\begin{split}
\widehat{F}_{\mathbf{p}\tau}&=\begin{pmatrix}
O & \widehat{\Phi}_{\mathbf{p}\tau} \\
\widehat{\Phi}^\dag_{\mathbf{p}\tau} & O
\end{pmatrix},\\
\widehat{\Phi}_{\mathbf{p}+}&=-i\begin{pmatrix}
\delta^{(+)}_\mathbf{p} & \widetilde{\gamma}_\mathbf{p} \\
\gamma^{(+)}_\mathbf{p} & \widetilde{\delta}_\mathbf{p}
\end{pmatrix},~~
\widehat{\Phi}_{\mathbf{p}-}=-i\begin{pmatrix}
\widetilde{\delta}_\mathbf{p} & \gamma^{(-)}_\mathbf{p} \\
\widetilde{\gamma}_\mathbf{p} & \delta^{(-)}_\mathbf{p}
\end{pmatrix}.
\end{split}
\end{equation*}

Consider singlet pairing. New operators are expressed in terms of old as follows
\begin{equation}\label{NewOperators1}
\begin{pmatrix}
  \alpha_{\mathbf{p}sK_+} \\
  \alpha_{\mathbf{p}sK_-} \\
  \beta^\dag_{-\mathbf{p}-sK_{-\sgn(s)}}
\end{pmatrix}=
M_{\sgn(s)}\begin{pmatrix}
  a_{\mathbf{p}sK_+} \\
  a_{\mathbf{p}sK_-} \\
  b^\dag_{-\mathbf{p}-sK_{-\sgn(s)}}
\end{pmatrix},
\end{equation}
where the rotation matrix $M_{\sgn(s)}$ depends on two angles (it is given in Appendix B). The angle $\phi^{(\sgn(s))}_\mathbf{p}$ determines the position of the rotation axis in the $xy$ plane (it lies at the angle $-\phi^{(\sgn(s))}_ \mathbf{p}$ to the $x$ axis, and $0<\phi^{(\sgn(s))}_\mathbf{p}<\hspace{-0.05cm}\,^\pi\hspace{-0.05cm}/_2$), and the angle $\varphi^{(\sgn(s))}_\mathbf{p}$ is the angle of rotation around it. These quantities are expressed in terms of the functions introduced by us as follows
\begin{equation}\label{Angles1}
\begin{split}
&\varphi^{(\pm)}_\mathbf{p}=\sqrt{\frac{\widetilde{\gamma}^2_\mathbf{p}+\gamma^{(\mp)2}_\mathbf{p}}{2}},\\
&\cos\phi^{(+)}_\mathbf{p}=\frac{\gamma^{(-)}_\mathbf{p}}{\sqrt{\widetilde{\gamma}^2_\mathbf{p}+\gamma^{(-)2}_\mathbf{p}}},\\
&\cos\phi^{(-)}_\mathbf{p}=\frac{\widetilde{\gamma}_\mathbf{p}}{\sqrt{\widetilde{\gamma}^2_\mathbf{p}+\gamma^{(+)2}_\mathbf{p}}}.
\end{split}
\end{equation}

For triplet pairing, when instead of $\beta^\dag_{-\mathbf{p}-sK_{-\sgn(s)}}$ and $b^\dag_{-\mathbf{p}-sK_{-\sgn (s)}}$ in \eqref{NewOperators1} cost $\beta^\dag_{-\mathbf{p}sK_{\sgn(s)}}$ and $b^\dag_{-\mathbf{p}sK_{\sgn(s)}}$, respectively, functions $\phi^{(\sgn(s))}_\mathbf{p}$ and $\varphi^{(\sgn(s))}_\mathbf{p}$ is chosen so that the matrix $M_{\sgn(s)}$ remains the same
\begin{equation}\label{Angles2}
\begin{split}
&\varphi^{(\pm)}_\mathbf{p}=\sqrt{\frac{\widetilde{\delta}^2_\mathbf{p}+\delta^{(\pm)2}_\mathbf{p}}{2}},\\
&\cos\phi^{(+)}_\mathbf{p}=\frac{\widetilde{\delta}_\mathbf{p}}{\sqrt{\widetilde{\delta}^2_\mathbf{p}+\delta^{(+)2}_\mathbf{p}}},\\
&\cos\phi^{(-)}_\mathbf{p}=\frac{\delta^{(-)}_\mathbf{p}}{\sqrt{\widetilde{\delta}^2_\mathbf{p}+\delta^{(-)2}_\mathbf{p}}}.
\end{split}
\end{equation}

The relations \eqref{Angles1} and \eqref{Angles2} reveal the mutual dependence of the angles $\varphi^{(\sgn(s))}_\mathbf{p}$ and $\phi^{(\sgn(s))} _\mathbf{p}$ for singlet and triplet pairing types, respectively.

The expressions \eqref{NewOperators} are generalized to the case of unequal population of valleys as follows
\begin{equation}\label{NewOperators2}
\begin{split}
&\alpha_{\mathbf{p}sK_\tau}=\frac{1}{2}\left(1+\tau\cos2\phi^{(\sgn(s))}_\mathbf{p}+\right.\\
&\left.+\left(1-\tau\cos2\phi^{(\sgn(s))}_\mathbf{p}\right)\cos\varphi^{(\sgn(s))}_\mathbf{p}\right)a_{\mathbf{p}sK_\tau}+\\
&+\frac{1}{2}\sin2\phi^{(\sgn(s))}_\mathbf{p}\left(\cos\varphi^{(\sgn(s))}_\mathbf{p}-1\right)a_{\mathbf{p}sK_{-\tau}}+\\
&+\sqrt{\frac{1-\tau\cos2\phi^{(\sgn(s))}_\mathbf{p}}{2}}\sin\varphi^{(\sgn(s))}_\mathbf{p}\times\\
&\times\left(\delta_{\widetilde{S}0}b^\dag_{-\mathbf{p}-sK_{-\sgn(s)}}+\delta_{\widetilde{S}1}b^\dag_{-\mathbf{p}sK_{\sgn(s)}}\right),\\
&\beta_{\mathbf{p}sK_{\sgn(s)}}=\cos\varphi^{(\sgn(s))}_\mathbf{p}b_{\mathbf{p}sK_{\sgn(s)}}-\\
&-\sin\varphi^{(\sgn(s))}_\mathbf{p}\sum_\tau\sqrt{\frac{1-\tau\cos2\phi^{(\sgn(s))}_\mathbf{p}}{2}}\times\\
&\times\left(\delta_{\widetilde{S}0}a^\dag_{-\mathbf{p}-sK_\tau}+\delta_{\widetilde{S}1}a^\dag_{-\mathbf{p}sK_\tau}\right).
\end{split}
\end{equation}
If we put $\phi^{(\sgn(s))}_\mathbf{p}\equiv\hspace{-0.05cm}\,^\pi\hspace{-0.05cm}/_4$ we return to the formulas \eqref{NewOperators}.

We note the feature of new quasiparticles, which follows from the formulas \eqref{NewOperators1}--\eqref{NewOperators2}. For different projections of the spin $s$, the pairing of an electron and a hole (both singlet and triplet) occurs differently --- they are described by different rotations. This is a reflection of the fact that the ensemble of old quasiparticles was initially partially spin-polarized (the number of holes with spin up is not equal to the number of holes with spin down with unequal population of valleys).

The transformed Hamiltonian $\widehat{\mathcal{H}}=S\widehat{H}S^\dag$ has the form
\begin{equation}\label{NewHam1}
\widehat{\mathcal{H}}=\widetilde{U}\left\{\varphi^{(\pm)}_\mathbf{p},\,\phi^{(\pm)}_\mathbf{p}\right\}+\widehat{\mathcal{H}}_0+\widehat{\mathcal{H}}_i-\mu n.
\end{equation}
The first term in \eqref{NewHam1} is a generalization of the numeric functional \eqref{Ut}
\begin{equation}\label{Ut1}
\begin{split}
&\widetilde{U}\left\{\varphi^{(\pm)}_\mathbf{p},\,\phi^{(\pm)}_\mathbf{p}\right\}=\sum_{\mathbf{p}s}\varepsilon_\mathbf{p}\sin^2\varphi^{(\sgn(s))}_\mathbf{p}-\\
&-\sum_{\mathbf{p}\mathbf{p}^\prime s}V_{\mathbf{p}-\mathbf{p}^\prime}\left[\frac{1}{2}\left(1+\cos^2\left(\phi^{(\sgn(s))}_\mathbf{p}-\phi^{(\sgn(s))}_{\mathbf{p}^\prime}\right)\right)\times\right.\\
&\times\sin^2\varphi^{(\sgn(s))}_\mathbf{p}\sin^2\varphi^{(\sgn(s))}_{\mathbf{p}^\prime}+\\
&+\cos\left(\phi^{(\sgn(s))}_\mathbf{p}-\phi^{(\sgn(s))}_{\mathbf{p}^\prime}\right)\times\\
&\left.\times\cos\varphi_\mathbf{p}\sin\varphi_\mathbf{p}\cos\varphi_{\mathbf{p}^\prime}\sin\varphi_{\mathbf{p}^\prime}\right].
\end{split}
\end{equation}

The operators $\widehat{\mathcal{H}}_0$ and $\widehat{\mathcal{H}}_i$ are given in Appendix C.

When the equalities \eqref{Norm_phi1} are satisfied, the condition \eqref{Norm_phi2} in the case of unequal population of the valleys takes the form
\begin{equation}\label{Norm_phi3}
\sum_{\mathbf{p}s}\sin^2\varphi^{(\sgn(s))}_\mathbf{p}=n.
\end{equation}

To take into account the relation between the charge carrier densities $n_+$ and $n_-$ belonging to the valleys at the points $K_+$ and $K_-$, respectively, it is necessary to rewrite \eqref{Norm_phi3} in more detail [according to the note $^{1}$]
\begin{equation}\label{Norm_phi4}
\sum_{\mathbf{p}s}\langle\alpha^\dag_{\mathbf{p}sK_\pm}\alpha_{\mathbf{p}sK_\pm}\rangle=\sum_\mathbf{p}\langle\beta^\dag_{\mathbf{p}\pm\frac{1}{2}K_\pm}\beta_{\mathbf{p}\pm\frac{1}{2}K_\pm}\rangle=n_\pm.
\end{equation}
After substituting \eqref{NewOperators2} into \eqref{Norm_phi4} we get the equalities
\begin{equation}\label{Norm_phi5}
\begin{split}
\sum_{\mathbf{p}s}\sin^2\phi^{(\sgn(s))}_\mathbf{p}\sin^2\varphi^{(\sgn(s))}_\mathbf{p}&=n_+,\\
\sum_{\mathbf{p}s}\cos^2\phi^{(\sgn(s))}_\mathbf{p}\sin^2\varphi^{(\sgn(s))}_\mathbf{p}&=n_-.
\end{split}
\end{equation}

Let us pass from summation over $\mathbf{p}$ to integration over $q=p/\widetilde{p}$. The value of $\widetilde{p}$ is determined by the relation \eqref{pt} with $p_0$, now equal to
\begin{equation*}
p_0=\frac{1}{\sqrt{\nu_e}}\sqrt{\int\limits_0^\infty\frac{qdq}{1+z^2_{q+}}+\int\limits_0^\infty\frac{qdq}{1+z^2_{q-}}}
\end{equation*}
with the functions $z_{q\pm}=\cot\varphi^{(\pm)}_q$.

The energy of an electron-hole pair is determined by minimizing the functional

\begin{equation}\label{Ut1}
\begin{split}
&E_0\left\{z_{q+},\,z_{q_-};\,\phi^{(+)}_q,\,\phi^{(-)}_q\right\}=\\
&=\frac{2}{\nu_er^2_sp^4_0}\sum_s\int\limits_0^\infty\frac{q^3dq}{1+z^2_{q\sgn(s)}}-\\
&-\frac{2\sqrt{2}}{\pi\nu_er_sp^3_0}\sum_s\int\limits_0^\infty\frac{q^2dq}{1+z^2_{q\sgn(s)}}\int\limits_0^1\frac{1}{1+z^2_{q\xi\sgn(s)}}\times\\
&\times\left(1+\cos^2\left(\phi^{(\sgn(s))}_q-\phi^{(\sgn(s))}_{q\xi}\right)+\right.\\
&\left.+2\cos\left(\phi^{(\sgn(s))}_q-\phi^{(\sgn(s))}_{q\xi}\right)z_{q\sgn(s)}z_{q\xi\sgn(s)}\right)\times\\
&\times\left[K(\xi)-\frac{1}{1+\xi}\widetilde{K}\left(\frac{2\sqrt{\xi}}{1+\xi};\,\widetilde{r}_0q(1+\xi)\right)\right]\xi d\xi.
\end{split}
\end{equation}

Trial functions $z_{q\pm}$ are chosen similarly to \eqref{TrialFunctions} with variational parameters $A_\pm$ and $B_\pm$ respectively. The functions $\phi^{(\pm)}_q$ are chosen according to the relations \eqref{Angles1} or \eqref{Angles2} [up to renaming, both relations lead to the same result]
\begin{equation}\label{phi}
\phi^{(\pm)}_q=\arccos\left(\frac{\arctan\left[A^{(\pm)}(1+q^2)^\alpha+B^{(\pm)}\right]^{-1}}{\sqrt{2}\arctan\left[A_\pm(1+q^2)^\alpha+B_\pm\right]^{-1}}\right).
\end{equation}
The assumption about the similarity of the functions $\widetilde{\gamma}_\mathbf{p}$ and $\gamma^{(\pm)}_\mathbf{p}$ ($\widetilde{\delta}_\mathbf {p}$ and $\delta^{(\pm)}_\mathbf{p}$).

The values of the parameters $A^{(\pm)}$ and $B^{(\pm)}$ are limited with respect to the values of the parameters $A_\pm$ and $B_\pm$ by the condition that the fraction does not exceed units in \eqref{phi}. They are also restricted by the conditions \eqref{Norm_phi5}.

Correlation corrections are calculated in the same way as for equal population of valleys. In view of their smallness, we consider such an approximation admissible.

\begin{center}
\textbf{V. TRANSITION BETWEEN METALLIC AND DIELECTRIC EHL}
\end{center}

As shown by numerical calculations (see Fig. 2), with equal population of the valleys, the metallic EHL in monolayer heterostructures based on TMD always turns out to be more energetically favorable than the dielectric EHL, which is consistent with our previous results \cite{Pekh2020, Pekh2021}. However, it can also be seen from \cite{Pekh2021} that there is a significant dependence of the binding energy of the metallic EHL on the number of valleys: for a single-valley semiconductor (under the condition of spin degeneracy of both electrons and holes), it exceeds the exciton binding energy by only 9\%. This suggests that a further decrease in the degeneracy multiplicity due to the removal of spin degeneracy for holes will lead to an even lower value of the binding energy of the metallic EHL.

Numerical calculations have shown that for a number of TMD monolayers and heterostructures based on them, when the degree of circular polarization of the exciting light $P_e$ tends to 1, the metallic EHL ceases to exist (its binding energy becomes lower than the exciton binding energy). In this case, the binding energy of the dielectric EHL is almost halved, but it still remains energetically favorable compared to the exciton (see Fig.~\hyperlink{f3}{3}).

\begin{figure}[b!]
\begin{center}
\hypertarget{f3}{}
\includegraphics[width=0.45\textwidth]{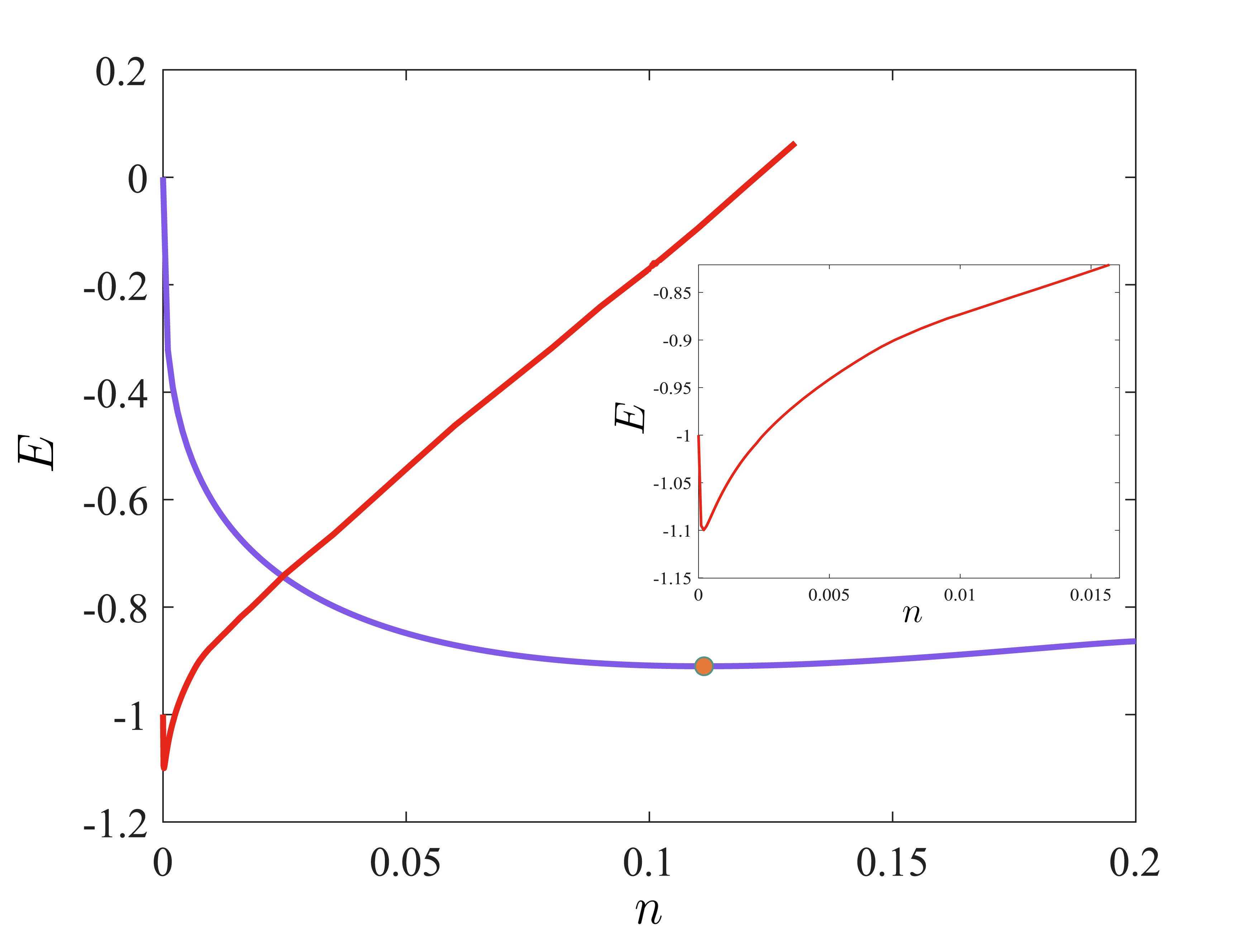}
\end{center}
FIG. 3. Dependence of the energy of the dielectric EHL (red curve) and metallic EHL (light blue curve) on the charge carrier density $n$ in the MoS$_2$ monolayer on the SiO$_2$ substrate in the case of one valley being populated. The yellow dot marks the minimum energy of the metallic EHL (its binding energy with a minus sign). The inset shows an enlarged section of the curve for the dielectric EHL.
\end{figure}

It follows that for some difference in the filling of the valleys, when $P_e$ lies between 0.5 and 1, the binding energies of both types of EHL are compared. This means that there is a threshold value of the degree of circular polarization of the exciting radiation $P_{e0}$, below which there is a metallic EHL, and above it, a dielectric EHL.

For most monolayers, the TMD $P_{e0}$ lies in the range 0.7--0.9. For example, for MoS$_2$ $P_{e0}=0.83$.

Two schemes can be used to experimentally observe a phase transition between two types of EHL. First scheme with two sources:
one is right polarized radiation, the other is left polarized. The change in the intensity of both sources is carried out synchronously so that the total intensity remains constant. The second option is to use a single source of linearly polarized light. The beam from it is divided into two beams, one of which is amplified with intensity modulation in time, and then added to the second (non-amplified). In this way it is possible to obtain radiation with $P_e$ modulated in time.

When $P_e$ exceeds the threshold value $P_{e0}$, droplets of the metallic EHL will disintegrate with the formation of a more ``loos'' phase, i.e. dielectric EHL. On the contrary, as $P_e$ decreases, when $P_e$ becomes less than $P_{e0}$, droplets of metal EHL will be formed again. Such ``walking'' of $P_e$ through $P_{e0}$ should lead, in particular, to the characteristic emission of phonons. A detailed development of this issue deserves a separate publication.

Finally, we note that, in contrast to the metallic EHL, a detailed calculation of the phase diagram of the gas-liquid transition in the case of a dielectric EHL loses its meaning, since its equilibrium density (at $T=0$) $n^{(d)}_0$ is hundreds of times less than that for the metal EHL. The bell-shaped curve of the coexistence of an exciton gas--EHL \cite{Pekh2021} ``compresses'' along the $n$ axis and ``shrinks'' along the $T$ axis, becoming similar to a finger-like curve in the corresponding number of times. The critical point of the gas-liquid transition is determined by well-established similarity relations \cite{Pekh2021, Andryushin1977}
\begin{equation}
T_c\simeq\frac{1}{10}|E_0|~~\text{and}~~n_c\simeq\frac{1}{5}n^{(d)}_0.
\end{equation}

In addition, at the excitation light intensities characteristic of the formation of a metallic EHL, the dielectric EHL will most likely fill the entire sample rather than form in the form of individual drops with an exciton gas between them. At sufficiently high densities, the coexistence of droplets of a metallic EHL and a dielectric EHL is possible: the droplets ``immersed'' in the dielectric EHL spilled over the rest of the sample area.

\begin{center}
\textbf{VI. R\'{E}SUM\'{E}}
\end{center}

In this work, we succeeded in showing that the dielectric EHL is possible in TMD monolayers and heterostructures based on them due to the coherent pairing of electrons and holes. An appropriately adapted canonical transformation was used to construct the correct zeroth approximation. Numerical calculations show that in the case of equal population of the valleys, the metallic EHL is more energetically favorable than the dielectric EHL. However, in the case of unequal population of the valleys, when the difference between the populations of the valleys is sufficiently large, the dielectric EHL becomes more energetically favorable. This occurs when the threshold value of the degree of circular polarization of the exciting light is exceeded. We also briefly describe the possible ways of experimental observation of the metallic EHL--dielectric EHL transition in addition to the standard detection method by analyzing the luminescence spectrum. We have pointed out the characteristic features of the corresponding phase diagram.

\begin{center}
\textbf{ACKNOWLEDGMENTS}
\end{center}

The author expresses his deep gratitude to S.~G.~Ti-khodeev for fruitful discussions of the results obtained. The work was supported by the Foundation for the Advancement of Theoretical Physics and Mathematics ``BASIS'' (project no. 20-1-3-68-1).

\onecolumn

\begin{center}
\textbf{APPENDIX A}
\end{center}

The operator $\widehat{\mathcal{H}}_0$ contains terms that are bilinear in Fermi operators (in contrast to \cite{Keldysh1968}, here it is taken into account that $m_e\neq m_h$, and the specifics of the band structure of TMD monolayers are taken into account)
\begin{equation*}
\begin{split}
&\widehat{\mathcal{H}}_0=\frac{1}{2}\sum_{\mathbf{p}s\tau}\left\{\left[\cos2\varphi_\mathbf{p}\left(\frac{1}{2}\xi_\mathbf{p}-\mathcal{V}_\mathbf{p}\right)+
\sin2\varphi_\mathbf{p}\widetilde{\mathcal{V}}_\mathbf{p}\right]\left(a^\dag_{\mathbf{p}sK_\tau}a_{\mathbf{p}sK_\tau}+b^\dag_{\mathbf{p}sK_{\sgn(s)}}b_{\mathbf{p}sK_{\sgn(s)}}\right)-\right.\\
&-\left[\xi_\mathbf{p}\sin^2\varphi_\mathbf{p}+\cos2\varphi_\mathbf{p}\mathcal{V}_\mathbf{p}-\sin2\varphi_\mathbf{p}\widetilde{\mathcal{V}}_\mathbf{p}\right]
a^\dag_{\mathbf{p}sK_\tau}a_{\mathbf{p}sK_{-\tau}}+\frac{1}{2}\delta\xi_\mathbf{p}\left(a^\dag_{\mathbf{p}sK_\tau}a_{\mathbf{p}sK_\tau}-b^\dag_{\mathbf{p}sK_{\sgn(s)}}b_{\mathbf{p}sK_{\sgn(s)}}\right)+\\
&\left.+\xi^e_\mathbf{p}a^\dag_{\mathbf{p}sK_\tau}a_{\mathbf{p}sK_\tau}\right\}+\frac{1}{\sqrt{2}}\sum_{\mathbf{p}s\tau}\left[\sin2\varphi_\mathbf{p}\left(\frac{1}{2}\xi_\mathbf{p}-\mathcal{V}_\mathbf{p}\right)-
\cos2\varphi_\mathbf{p}\widetilde{\mathcal{V}}_\mathbf{p}\right]\left(a^\dag_{\mathbf{p}sK_\tau}b^\dag_{-\mathbf{p}-sK_{-\sgn(s)}}+b_{-\mathbf{p}-sK_{-\sgn(s)}}a_{\mathbf{p}sK_\tau}\right),
\end{split}
\end{equation*}
where the notation is introduced: $\xi_\mathbf{p}=\varepsilon_\mathbf{p}-\mu$, $\xi^{e,h}_\mathbf{p}=\varepsilon^{e,h}_\mathbf{p}-\mu_{e,h}$ ($\xi_\mathbf{p}\equiv\xi^e_\mathbf{p}+\xi^h_\mathbf{p}$), $\delta\xi_\mathbf{p}=\xi^e_\mathbf{p}-\xi^h_\mathbf{p}$,
\begin{equation*}
\mathcal{V}_\mathbf{p}=\sum_{\mathbf{p}^\prime}V_{\mathbf{p}-\mathbf{p}^\prime}\sin^2\varphi_{\mathbf{p}^\prime}~~\text{and}~~
\widetilde{\mathcal{V}}_\mathbf{p}=\sum_{\mathbf{p}^\prime}V_{\mathbf{p}-\mathbf{p}^\prime}\cos\varphi_{\mathbf{p}^\prime}\sin\varphi_{\mathbf{p}^\prime}.
\end{equation*}

The operator $\widehat{\mathcal{H}}_i$ contains four combinations of Fermi operators
\begin{equation*}
\begin{split}
\widehat{\mathcal{H}}_i=\frac{1}{4}\sum_{\substack{\mathbf{p}\mathbf{p}^\prime\mathbf{k}\\ ss^\prime\tau\tau^\prime}}
V_\mathbf{k}&\left\{\frac{1}{2}\left(\gamma_{\mathbf{p},\,\mathbf{p}-\mathbf{k}}+1\right)\left(\gamma_{\mathbf{p}^\prime,\,\mathbf{p}^\prime+\mathbf{k}}+1\right)
a^\dag_{\mathbf{p}sK_\tau}a^\dag_{\mathbf{p}^\prime s^\prime K_{\tau^\prime}}a_{\mathbf{p}^\prime+\mathbf{k}s^\prime K_{\tau^\prime}}a_{\mathbf{p}-\mathbf{k}sK_\tau}+\right.\\
&+\frac{1}{2}\left(\gamma_{\mathbf{p},\,\mathbf{p}-\mathbf{k}}-1\right)\left(\gamma_{\mathbf{p}^\prime,\,\mathbf{p}^\prime+\mathbf{k}}+1\right)
a^\dag_{\mathbf{p}sK_\tau}a^\dag_{\mathbf{p}^\prime s^\prime K_{\tau^\prime}}a_{\mathbf{p}^\prime+\mathbf{k}s^\prime K_{\tau^\prime}}a_{\mathbf{p}-\mathbf{k}sK_{-\tau}}+\\
&+\frac{1}{2}\left(\gamma_{\mathbf{p},\,\mathbf{p}-\mathbf{k}}+1\right)\left(\gamma_{\mathbf{p}^\prime,\,\mathbf{p}^\prime+\mathbf{k}}-1\right)
a^\dag_{\mathbf{p}sK_\tau}a^\dag_{\mathbf{p}^\prime s^\prime K_{\tau^\prime}}a_{\mathbf{p}^\prime+\mathbf{k}s^\prime K_{-\tau^\prime}}a_{\mathbf{p}-\mathbf{k}sK_\tau}+\\
&+\frac{1}{2}\left(\gamma_{\mathbf{p},\,\mathbf{p}-\mathbf{k}}-1\right)\left(\gamma_{\mathbf{p}^\prime,\,\mathbf{p}^\prime+\mathbf{k}}-1\right)
a^\dag_{\mathbf{p}sK_\tau}a^\dag_{\mathbf{p}^\prime s^\prime K_{\tau^\prime}}a_{\mathbf{p}^\prime+\mathbf{k}s^\prime K_{-\tau^\prime}}a_{\mathbf{p}-\mathbf{k}sK_{-\tau}}+\\
&+\frac{1}{2}\gamma_{\mathbf{p},\,\mathbf{p}-\mathbf{k}}\gamma_{\mathbf{p}^\prime,\,\mathbf{p}^\prime+\mathbf{k}}b^\dag_{\mathbf{p}sK_{\sgn(s)}}b^\dag_{\mathbf{p}^\prime s^\prime K_{\sgn(s^\prime)}}b_{\mathbf{p}^\prime+\mathbf{k}s^\prime K_{\sgn(s^\prime)}}b_{\mathbf{p}-\mathbf{k}sK_{\sgn(s)}}-\\
&-\left(\gamma_{\mathbf{p},\,\mathbf{p}-\mathbf{k}}+1\right)\gamma_{\mathbf{p}^\prime,\,\mathbf{p}^\prime+\mathbf{k}}a^\dag_{\mathbf{p}sK_\tau}b^\dag_{\mathbf{p}^\prime s^\prime K_{\sgn(s^\prime)}}b_{\mathbf{p}^\prime+\mathbf{k}s^\prime K_{\sgn(s^\prime)}}a_{\mathbf{p}-\mathbf{k}sK_\tau}-\\
&-\left(\gamma_{\mathbf{p},\,\mathbf{p}-\mathbf{k}}-1\right)\gamma_{\mathbf{p}^\prime,\,\mathbf{p}^\prime+\mathbf{k}}a^\dag_{\mathbf{p}sK_\tau}b^\dag_{\mathbf{p}^\prime s^\prime K_{\sgn(s^\prime)}}b_{\mathbf{p}^\prime+\mathbf{k}s^\prime K_{\sgn(s^\prime)}}a_{\mathbf{p}-\mathbf{k}sK_{-\tau}}+\\
&+\sqrt{2}\left(\gamma_{\mathbf{p},\,\mathbf{p}-\mathbf{k}}+1\right)\widetilde{\gamma}_{\mathbf{p}^\prime,\,\mathbf{p}^\prime+\mathbf{k}}
\left[a^\dag_{\mathbf{p}sK_\tau}a^\dag_{\mathbf{p}^\prime s^\prime K_{\tau^\prime}}b^\dag_{-\mathbf{p}^\prime-\mathbf{k}-s^\prime K_{-\sgn(s^\prime)}}a_{\mathbf{p}-\mathbf{k}sK_\tau}+\text{h.c.}\right]+\\
&+\sqrt{2}\left(\gamma_{\mathbf{p},\,\mathbf{p}-\mathbf{k}}-1\right)\widetilde{\gamma}_{\mathbf{p}^\prime,\,\mathbf{p}^\prime+\mathbf{k}}
\left[a^\dag_{\mathbf{p}sK_\tau}a^\dag_{\mathbf{p}^\prime s^\prime K_{\tau^\prime}}b^\dag_{-\mathbf{p}^\prime-\mathbf{k}-s^\prime K_{-\sgn(s^\prime)}}a_{\mathbf{p}-\mathbf{k}sK_{-\tau}}+\text{h.c.}\right]-\\
&-\sqrt{2}\gamma_{\mathbf{p},\,\mathbf{p}-\mathbf{k}}\widetilde{\gamma}_{\mathbf{p}^\prime,\,\mathbf{p}^\prime+\mathbf{k}}
\left[a^\dag_{\mathbf{p}^\prime s^\prime K_{\tau^\prime}}b^\dag_{-\mathbf{p}^\prime-\mathbf{k}-s^\prime K_{-\sgn(s^\prime)}}b^\dag_{\mathbf{p}s K_{\sgn(s)}}b_{\mathbf{p}-\mathbf{k}sK_{\sgn(s)}}+\text{h.c.}\right]+\\
&+\widetilde{\gamma}_{\mathbf{p},\,\mathbf{p}-\mathbf{k}}\widetilde{\gamma}_{\mathbf{p}^\prime,\,\mathbf{p}^\prime+\mathbf{k}}
\left[a^\dag_{\mathbf{p}sK_\tau}a^\dag_{\mathbf{p}^\prime s^\prime K_{\tau^\prime}}b^\dag_{-\mathbf{p}^\prime-\mathbf{k}-s^\prime K_{-\sgn(s^\prime)}}b^\dag_{-\mathbf{p}+\mathbf{k}-sK_{-\sgn(s)}}+\right.\\
&\left.\left.+a^\dag_{\mathbf{p}sK_\tau}b^\dag_{-\mathbf{p}+\mathbf{k}-sK_{-\sgn(s)}}b_{\mathbf{p}^\prime+\mathbf{k}-s^\prime K_{-\sgn(s^\prime)}}a_{-\mathbf{p}^\prime s^\prime K_{\tau^\prime}}+\text{h.c.}\right]\right\}.
\end{split}
\end{equation*}
Here, as in \cite{Keldysh1968}, the functions
\begin{equation*}
\gamma_{\mathbf{p},\,\mathbf{p}^\prime}=\cos\left(\varphi_\mathbf{p}-\varphi_{\mathbf{p}^\prime}\right)~~\text{and}~~
\widetilde{\gamma}_{\mathbf{p},\,\mathbf{p}^\prime}=\sin\left(\varphi_{\mathbf{p}^\prime}-\varphi_\mathbf{p}\right).
\end{equation*}
Considering that $\sin\varphi_\mathbf{p}\sim\sqrt{n}$ and $\cos\varphi_\mathbf{p}\sim1-\mathcal{O}(n)$ (according to the condition \eqref{Norm_phi2}), find $\gamma_{\mathbf{p},\,\mathbf{p}^\prime}\sim1-\mathcal{O}(n)$ and $\widetilde{\gamma}_{\mathbf {p},\,\mathbf{p}^\prime}\sim\sqrt{n}$ \cite{Keldysh1968}. From this we obtain that the terms responsible for ``transfer'' of an electron from one valley to another are suppressed as $\gamma_{\mathbf{p},\,\mathbf{p}-\mathbf{k}}-1\sim n$, and the terms with double intervalley ``transfer'' electrons are suppressed as $\left(\gamma_{\mathbf{p},\,\mathbf{p}-\mathbf{k}}-1\right)\left(\gamma_{\mathbf{p}^\prime,\,\mathbf{p}^\prime+\mathbf{k}}-1\right)\sim n^2$.

~

\textbf{APPENDIX B}

The rotation matrix in \eqref{NewOperators1} is
\begin{equation*}
\begin{split}
&M_{\sgn(s)}\equiv M_\pm\left(\varphi^{(\pm)}_\mathbf{p},\,\phi^{(\pm)}_\mathbf{p}\right)=\\
&=\begin{pmatrix}
    \cos^2\phi^{(\pm)}_\mathbf{p}+\sin^2\phi^{(\pm)}_\mathbf{p}\cos\varphi^{(\pm)}_\mathbf{p} & \cos\phi^{(\pm)}_\mathbf{p}\sin\phi^{(\pm)}_\mathbf{p}\left(\cos\varphi^{(\pm)}_\mathbf{p}-1\right) & \sin\phi^{(\pm)}_\mathbf{p}\sin\varphi^{(\pm)}_\mathbf{p} \\
    \cos\phi^{(\pm)}_\mathbf{p}\sin\phi^{(\pm)}_\mathbf{p}\left(\cos\varphi^{(\pm)}_\mathbf{p}-1\right) & \sin^2\phi^{(\pm)}_\mathbf{p}+\sin^2\phi^{(\pm)}_\mathbf{p}\cos\varphi^{(\pm)}_\mathbf{p} & \cos\phi^{(\pm)}_\mathbf{p}\sin\varphi^{(\pm)}_\mathbf{p} \\
    -\sin\phi^{(\pm)}_\mathbf{p}\sin\varphi^{(\pm)}_\mathbf{p} & -\cos\phi^{(\pm)}_\mathbf{p}\sin\varphi^{(\pm)}_\mathbf{p} & \cos\varphi^{(\pm)}_\mathbf{p}
  \end{pmatrix}.
\end{split}
\end{equation*}
Here, as in the main text of the article, $\pm$ is the same as $\sgn(s)$.

~

\textbf{APPENDIX C}

The operator $\widehat{\mathcal{H}}_0$ in the expression \eqref{NewHam1} is
\begin{equation*}
\begin{split}
&\widehat{\mathcal{H}}_0=\frac{1}{2}\sum_{\mathbf{p}s\tau}\left\{\left(1-\tau\cos2\phi^{(\sgn(s))}_\mathbf{p}\right)
\left(\frac{1}{2}\xi_\mathbf{p}-\mathcal{V}^{(1)}_{\mathbf{p}s}\right)\cos^2\varphi^{(\sgn(s))}_\mathbf{p}-\sqrt{1-\tau\cos2\phi^{(\sgn(s))}_\mathbf{p}}
\left(\sqrt{1-\tau\cos2\phi^{(\sgn(s))}_\mathbf{p}}\times\right.\right.\\
&\left.\times\frac{1}{2}\xi_\mathbf{p}-\mathcal{V}_{\mathbf{p}s\tau}\right)\sin^2\varphi^{(\sgn(s))}_\mathbf{p}
-\left(1+\tau\cos2\phi^{(\sgn(s))}_\mathbf{p}\right)\mathcal{V}^{(2)}_{\mathbf{p}s}+\tau\sin2\phi^{(\sgn(s))}_\mathbf{p}\cos\varphi^{(\sgn(s))}_\mathbf{p}\mathcal{V}^{(3)}_{\mathbf{p}s}+\\
&+\left(1-\tau\cos2\phi^{(\sgn(s))}_\mathbf{p}\right)\sin2\varphi^{(\sgn(s))}_\mathbf{p}\mathcal{V}^{(4)}_{\mathbf{p}s}-4\tau\sin2\phi^{(\sgn(s))}_\mathbf{p}\sin\varphi^{(\sgn(s))}_\mathbf{p}\mathcal{V}^{(5)}_{\mathbf{p}s}+
\frac{1}{2}\left(1-\tau\cos2\phi^{(\sgn(s))}_\mathbf{p}\right)\delta\xi_\mathbf{p}+\\
&\left.+\left(1+\tau\cos2\phi^{(\sgn(s))}_\mathbf{p}\right)\xi^e\right\}a^\dag_{\mathbf{p}sK_\tau}a_{\mathbf{p}sK_\tau}+
\frac{1}{2}\sum_{\mathbf{p}s\tau}\left\{-\sin2\phi^{(\sgn(s))}_\mathbf{p}\left[\xi_\mathbf{p}\sin^2\varphi^{(\sgn(s))}_\mathbf{p}+\cos^2\varphi^{(\sgn(s))}_\mathbf{p}\mathcal{V}^{(1)}_{\mathbf{p}s}-\right.\right.\\
&\left.-\mathcal{V}^{(2)}_{\mathbf{p}s}-\sin2\phi^{(\sgn(s))}_\mathbf{p}\mathcal{V}^{(4)}_{\mathbf{p}s}\right]+\sqrt{1-\tau\cos2\phi^{(\sgn(s))}_\mathbf{p}}\sin^2\varphi^{(\sgn(s))}_\mathbf{p}\mathcal{V}_{\mathbf{p}s-\tau}
+\cos2\phi^{(\sgn(s))}_\mathbf{p}\left(\cos\varphi^{(\sgn(s))}_\mathbf{p}\mathcal{V}^{(3)}_{\mathbf{p}s}\right.-\\
&\left.\left.-4\sin\varphi^{(\sgn(s))}_\mathbf{p}\mathcal{V}^{(5)}_{\mathbf{p}s}\right)\right\}a^\dag_{\mathbf{p}sK_\tau}a_{\mathbf{p}sK_{-\tau}}+
\sum_{\mathbf{p}s}\left\{\frac{1}{2}\xi_\mathbf{p}-\mathcal{V}^{(1)}_{\mathbf{p}s}-\cos^2\varphi^{(\sgn(s))}_\mathbf{p}\mathcal{V}^{(2)}_{\mathbf{p}s}+
\frac{1}{\sqrt{2}}\left(\sin\phi^{(\sgn(s))}_\mathbf{p}\widetilde{\mathcal{V}}_{\mathbf{p}s+}+\right.\right.\\
&\left.\left.+\cos\phi^{(\sgn(s))}_\mathbf{p}\widetilde{\mathcal{V}}_{\mathbf{p}s-}\right)\sin2\varphi^{(\sgn(s))}_\mathbf{p}-
\frac{1}{2}\delta\xi_\mathbf{p}\right\}b^\dag_{\mathbf{p}-sK_{-\sgn(s)}}b_{\mathbf{p}-sK_{-\sgn(s)}}+\frac{1}{\sqrt{2}}\sum_{\mathbf{p}s\tau}
\left[\sqrt{1-\tau\cos2\phi^{(\sgn(s))}_\mathbf{p}}\times\right.\\
&\times\left(\frac{1}{2}\xi_\mathbf{p}-\mathcal{V}^{(1)}_{\mathbf{p}s}-\frac{1}{2}\mathcal{V}^{(2)}_{\mathbf{p}s}\right)\sin2\varphi^{(\sgn(s))}_\mathbf{p}+
\frac{1}{2}\tau\sqrt{1+\tau\cos2\phi^{(\sgn(s))}_\mathbf{p}}\sin\varphi^{(\sgn(s))}_\mathbf{p}\mathcal{V}^{(3)}_{\mathbf{p}s}-\sqrt{1-\tau\cos2\phi^{(\sgn(s))}_\mathbf{p}}\times\\
&\left.\times\cos2\varphi^{(\sgn(s))}_\mathbf{p}\mathcal{V}^{(4)}_{\mathbf{p}s}+\tau\sqrt{1+\tau\cos2\phi^{(\sgn(s))}_\mathbf{p}}\cos\varphi^{(\sgn(s))}_\mathbf{p}\mathcal{V}^{(5)}_{\mathbf{p}s}\right]
\left(a^\dag_{\mathbf{p}sK_\tau}b^\dag_{-\mathbf{p}-sK_{-\sgn(s)}}+b_{-\mathbf{p}-sK_{-\sgn(s)}}a_{\mathbf{p}sK_\tau}\right),
\end{split}
\end{equation*}
where we introduced the functions
\begin{equation*}
\begin{split}
\mathcal{V}^{(1)}_{\mathbf{p}s}&=\sum_{\mathbf{p}^\prime}V_{\mathbf{p}-\mathbf{p}^\prime}\cos^2\left(\phi^{(\sgn(s))}_\mathbf{p}-\phi^{(\sgn(s))}_{\mathbf{p}^\prime}\right)\sin^2\varphi^{(\sgn(s))}_{\mathbf{p}^\prime},\\
\mathcal{V}^{(2)}_{\mathbf{p}s}&=\sum_{\mathbf{p}^\prime}V_{\mathbf{p}-\mathbf{p}^\prime}\sin^2\left(\phi^{(\sgn(s))}_\mathbf{p}-\phi^{(\sgn(s))}_{\mathbf{p}^\prime}\right)\sin^2\varphi^{(\sgn(s))}_{\mathbf{p}^\prime},\\
\mathcal{V}^{(3)}_{\mathbf{p}s}&=\sum_{\mathbf{p}^\prime}V_{\mathbf{p}-\mathbf{p}^\prime}\sin2\left(\phi^{(\sgn(s))}_\mathbf{p}-\phi^{(\sgn(s))}_{\mathbf{p}^\prime}\right)\sin^2\varphi^{(\sgn(s))}_{\mathbf{p}^\prime},\\
\mathcal{V}^{(4)}_{\mathbf{p}s}&=\sum_{\mathbf{p}^\prime}V_{\mathbf{p}-\mathbf{p}^\prime}\cos\left(\phi^{(\sgn(s))}_\mathbf{p}-\phi^{(\sgn(s))}_{\mathbf{p}^\prime}\right)\cos\varphi^{(\sgn(s))}_{\mathbf{p}^\prime}\sin\varphi^{(\sgn(s))}_{\mathbf{p}^\prime},\\
\mathcal{V}^{(5)}_{\mathbf{p}s}&=\sum_{\mathbf{p}^\prime}V_{\mathbf{p}-\mathbf{p}^\prime}\sin\left(\phi^{(\sgn(s))}_\mathbf{p}-\phi^{(\sgn(s))}_{\mathbf{p}^\prime}\right)\cos\varphi^{(\sgn(s))}_{\mathbf{p}^\prime}\sin\varphi^{(\sgn(s))}_{\mathbf{p}^\prime},\\
\mathcal{V}_{\mathbf{p}s\tau}&=\sum_{\mathbf{p}^\prime}V_{\mathbf{p}-\mathbf{p}^\prime}\sqrt{1-\tau\cos2\phi^{(\sgn(s))}_\mathbf{p}}\sin^2\varphi^{(\sgn(s))}_{\mathbf{p}^\prime},\\
\widetilde{\mathcal{V}}_{\mathbf{p}s\tau}&=\sum_{\mathbf{p}^\prime}V_{\mathbf{p}-\mathbf{p}^\prime}\sqrt{1-\tau\cos2\phi^{(\sgn(s))}_\mathbf{p}}\cos\varphi^{(\sgn(s))}_{\mathbf{p}^\prime}\sin\varphi^{(\sgn(s))}_{\mathbf{p}^\prime}.
\end{split}
\end{equation*}

The operator $\widehat{\mathcal{H}}_i$ in the expression \eqref{NewHam1} is
\begin{equation*}
\begin{split}
\widehat{\mathcal{H}}_i=\frac{1}{4}\sum_{\substack{\mathbf{p}\mathbf{p}^\prime\mathbf{k}\\ss^\prime\tau\tau^\prime}}V_\mathbf{k}&\left\{\frac{1}{2}
\mathcal{F}^{(+)}_{s\tau}\left(\mathbf{p},\,\mathbf{p}-\mathbf{k}\right)\mathcal{F}^{(+)}_{s^\prime\tau^\prime}\left(\mathbf{p}^\prime,\,\mathbf{p}^\prime+\mathbf{k}\right)
a^\dag_{\mathbf{p}sK_\tau}a^\dag_{\mathbf{p}^\prime s^\prime K_{\tau^\prime}}a_{\mathbf{p}^\prime+\mathbf{k}s^\prime K_{\tau^\prime}}a_{\mathbf{p}-\mathbf{k}sK_\tau}+\right.\\
&+\frac{1}{2}\mathcal{F}^{(+)}_{s\tau}\left(\mathbf{p},\,\mathbf{p}-\mathbf{k}\right)\mathcal{F}^{(-)}_{s^\prime\tau^\prime}\left(\mathbf{p}^\prime,\,\mathbf{p}^\prime+\mathbf{k}\right)
a^\dag_{\mathbf{p}sK_\tau}a^\dag_{\mathbf{p}^\prime s^\prime K_{\tau^\prime}}a_{\mathbf{p}^\prime+\mathbf{k}s^\prime K_{-\tau^\prime}}a_{\mathbf{p}-\mathbf{k}sK_\tau}+\\
&+\frac{1}{2}\mathcal{F}^{(-)}_{s\tau}\left(\mathbf{p},\,\mathbf{p}-\mathbf{k}\right)\mathcal{F}^{(+)}_{s^\prime\tau^\prime}\left(\mathbf{p}^\prime,\,\mathbf{p}^\prime+\mathbf{k}\right)
a^\dag_{\mathbf{p}sK_\tau}a^\dag_{\mathbf{p}^\prime s^\prime K_{\tau^\prime}}a_{\mathbf{p}^\prime+\mathbf{k}s^\prime K_{\tau^\prime}}a_{\mathbf{p}-\mathbf{k}sK_{-\tau}}+\\
&+\frac{1}{2}\mathcal{F}^{(-)}_{s\tau}\left(\mathbf{p},\,\mathbf{p}-\mathbf{k}\right)\mathcal{F}^{(-)}_{s^\prime\tau^\prime}\left(\mathbf{p}^\prime,\,\mathbf{p}^\prime+\mathbf{k}\right)
a^\dag_{\mathbf{p}sK_\tau}a^\dag_{\mathbf{p}^\prime s^\prime K_{\tau^\prime}}a_{\mathbf{p}^\prime+\mathbf{k}s^\prime K_{-\tau^\prime}}a_{\mathbf{p}-\mathbf{k}sK_{-\tau}}+\\
&+\frac{1}{2}\gamma_{\mathbf{p},\,\mathbf{p}-\mathbf{k},\,s}\gamma_{\mathbf{p}^\prime,\,\mathbf{p}^\prime+\mathbf{k},\,s}b^\dag_{\mathbf{p}-sK_{-\sgn(s)}}b^\dag_{\mathbf{p}^\prime -s^\prime K_{-\sgn(s^\prime)}}b_{\mathbf{p}^\prime+\mathbf{k}-s^\prime K_{-\sgn(s^\prime)}}b_{\mathbf{p}-\mathbf{k}-sK_{-\sgn(s)}}-\\
&-\mathcal{F}^{(+)}_{s\tau}\left(\mathbf{p},\,\mathbf{p}-\mathbf{k}\right)\gamma_{\mathbf{p}^\prime,\,\mathbf{p}^\prime+\mathbf{k},\,s^\prime}a^\dag_{\mathbf{p}sK_\tau}b^\dag_{\mathbf{p}^\prime -s^\prime K_{-\sgn(s^\prime)}}b_{\mathbf{p}^\prime+\mathbf{k}-s^\prime K_{-\sgn(s^\prime)}}a_{\mathbf{p}-\mathbf{k}sK_\tau}-\\
&-\mathcal{F}^{(-)}_{s\tau}\left(\mathbf{p},\,\mathbf{p}-\mathbf{k}\right)\gamma_{\mathbf{p}^\prime,\,\mathbf{p}^\prime+\mathbf{k},\,s^\prime}a^\dag_{\mathbf{p}sK_\tau}b^\dag_{\mathbf{p}^\prime -s^\prime K_{-\sgn(s^\prime)}}b_{\mathbf{p}^\prime+\mathbf{k}-s^\prime K_{-\sgn(s^\prime)}}a_{\mathbf{p}-\mathbf{k}sK_{-\tau}}+\\
&+\sqrt{2}\mathcal{F}^{(+)}_{s\tau}\left(\mathbf{p},\,\mathbf{p}-\mathbf{k}\right)\widetilde{\gamma}_{\mathbf{p}^\prime,\,\mathbf{p}^\prime+\mathbf{k},\,s^\prime,\,\tau^\prime}
\left[a^\dag_{\mathbf{p}sK_\tau}a^\dag_{\mathbf{p}^\prime s^\prime K_{\tau^\prime}}b^\dag_{-\mathbf{p}^\prime-\mathbf{k}-s^\prime K_{-\sgn(s^\prime)}}a_{\mathbf{p}-\mathbf{k}sK_\tau}+\text{h.c.}\right]+\\
&+\sqrt{2}\mathcal{F}^{(-)}_{s\tau}\left(\mathbf{p},\,\mathbf{p}-\mathbf{k}\right)\widetilde{\gamma}_{\mathbf{p}^\prime,\,\mathbf{p}^\prime+\mathbf{k},\,s^\prime,\,\tau^\prime}
\left[a^\dag_{\mathbf{p}sK_\tau}a^\dag_{\mathbf{p}^\prime s^\prime K_{\tau^\prime}}b^\dag_{-\mathbf{p}^\prime-\mathbf{k}-s^\prime K_{-\sgn(s^\prime)}}a_{\mathbf{p}-\mathbf{k}sK_{-\tau}}+\text{h.c.}\right]-\\
&-\sqrt{2}\gamma_{\mathbf{p},\,\mathbf{p}-\mathbf{k},\,s}\widetilde{\gamma}_{\mathbf{p}^\prime,\,\mathbf{p}^\prime+\mathbf{k},\,s^\prime,\,\tau^\prime}
\left[a^\dag_{\mathbf{p}^\prime s^\prime K_{\tau^\prime}}b^\dag_{-\mathbf{p}^\prime-\mathbf{k}-s^\prime K_{-\sgn(s^\prime)}}b^\dag_{\mathbf{p}-s K_{-\sgn(s)}}b_{\mathbf{p}-\mathbf{k}-sK_{-\sgn(s)}}+\text{h.c.}\right]+\\
&+\widetilde{\gamma}_{\mathbf{p},\,\mathbf{p}-\mathbf{k},\,s,\,\tau}\widetilde{\gamma}_{\mathbf{p}^\prime,\,\mathbf{p}^\prime+\mathbf{k},\,s^\prime,\,\tau^\prime}
\left[a^\dag_{\mathbf{p}sK_\tau}a^\dag_{\mathbf{p}^\prime s^\prime K_{\tau^\prime}}b^\dag_{-\mathbf{p}^\prime-\mathbf{k}-s^\prime K_{-\sgn(s^\prime)}}b^\dag_{-\mathbf{p}+\mathbf{k}-sK_{-\sgn(s)}}+\right.\\
&\left.\left.+a^\dag_{\mathbf{p}sK_\tau}b^\dag_{-\mathbf{p}+\mathbf{k}-sK_{-\sgn(s)}}b_{\mathbf{p}^\prime+\mathbf{k}-s^\prime K_{-\sgn(s^\prime)}}a_{-\mathbf{p}^\prime s^\prime K_{\tau^\prime}}+\text{h.c.}\right]\right\},
\end{split}
\end{equation*}
where we introduced the functions
\begin{equation*}
\begin{split}
&\mathcal{F}^{(\pm)}_{s\tau}\left(\mathbf{p},\,\mathbf{p}^\prime\right)=\sqrt{1-\tau\cos2\phi^{(\sgn(s))}_\mathbf{p}}\sqrt{1\mp\tau\cos2\phi^{(\sgn(s))}_{\mathbf{p}-\mathbf{k}}}\gamma^\prime_{\mathbf{p},\,\mathbf{p}^\prime,\,s}\pm\\
&\pm\sqrt{1+\tau\cos2\phi^{(\sgn(s))}_\mathbf{p}}\sqrt{1\pm\tau\cos2\phi^{(\sgn(s))}_{\mathbf{p}^\prime}}\cos\left(\phi^{(\sgn(s))}_\mathbf{p}-\phi^{(\sgn(s))}_{\mathbf{p}^\prime}\right)\pm\\
&\pm\tau\sin\left(\phi^{(\sgn(s))}_\mathbf{p}-\phi^{(\sgn(s))}_{\mathbf{p}^\prime}\right)\left[\sqrt{1-\tau\cos2\phi^{(\sgn(s))}_\mathbf{p}}\sqrt{1\pm\tau\cos2\phi^{(\sgn(s))}_{\mathbf{p}^\prime}}\cos\varphi^{(\sgn(s))}_\mathbf{p}\mp\right.\\
&\left.\mp\sqrt{1+\tau\cos2\phi^{(\sgn(s))}_\mathbf{p}}\sqrt{1\mp\tau\cos2\phi^{(\sgn(s))}_{\mathbf{p}-\mathbf{k}}}\cos\varphi^{(\sgn(s))}_{\mathbf{p}^\prime}\right],
\end{split}
\end{equation*}
\begin{equation*}
\begin{split}
&\gamma_{\mathbf{p},\,\mathbf{p}^\prime,\,s}=\cos\left(\phi^{(\sgn(s))}_\mathbf{p}-\phi^{(\sgn(s))}_{\mathbf{p}^\prime}\right)\sin\varphi^{(\sgn(s))}_\mathbf{p}\sin\varphi^{(\sgn(s))}_{\mathbf{p}^\prime}+\cos\varphi^{(\sgn(s))}_\mathbf{p}\cos\varphi^{(\sgn(s))}_{\mathbf{p}^\prime},\\
&\gamma^\prime_{\mathbf{p},\,\mathbf{p}^\prime,\,s}=\sin\varphi^{(\sgn(s))}_\mathbf{p}\sin\varphi^{(\sgn(s))}_{\mathbf{p}^\prime}+\cos\left(\phi^{(\sgn(s))}_\mathbf{p}-\phi^{(\sgn(s))}_{\mathbf{p}^\prime}\right)\cos\varphi^{(\sgn(s))}_\mathbf{p}\cos\varphi^{(\sgn(s))}_{\mathbf{p}^\prime},\\
&\widetilde{\gamma}_{\mathbf{p},\,\mathbf{p}^\prime,\,s,\,\tau}=\sqrt{1-\tau\cos2\phi^{(\sgn(s))}_\mathbf{p}}\widetilde{\gamma}_{\mathbf{p},\,\mathbf{p}^\prime,\,s}-
\tau\sqrt{1+\tau\cos2\phi^{(\sgn(s))}_\mathbf{p}}\sin\left(\phi^{(\sgn(s))}_\mathbf{p}-\phi^{(\sgn(s))}_{\mathbf{p}^\prime}\right)\sin\varphi^{(\sgn(s))}_{\mathbf{p}^\prime},\\
&\widetilde{\gamma}_{\mathbf{p},\,\mathbf{p}^\prime,\,s}=\cos\left(\phi^{(\sgn(s))}_\mathbf{p}-\phi^{(\sgn(s))}_{\mathbf{p}^\prime}\right)\cos\varphi^{(\sgn(s))}_\mathbf{p}\sin\varphi^{(\sgn(s))}_{\mathbf{p}^\prime}-\sin\varphi^{(\sgn(s))}_\mathbf{p}\cos\varphi^{(\sgn(s))}_{\mathbf{p}^\prime}.
\end{split}
\end{equation*}

%\newpage

\twocolumn

\end{document}